\documentclass{article}
\usepackage[utf8]{inputenc}

\usepackage[preprint]{neurips_2026}

\usepackage[utf8]{inputenc} 
\usepackage[T1]{fontenc}    
\usepackage{hyperref}       
\usepackage{url}            
\usepackage{booktabs}       
\usepackage{amsfonts}       
\usepackage{nicefrac}       
\usepackage{microtype}      
\usepackage{xcolor}         
\usepackage{graphicx}
\usepackage{amssymb} 
\usepackage{amsmath}
\usepackage{booktabs}
\usepackage{array}
\usepackage{enumitem}
\usepackage{algorithm}
\usepackage{wrapfig}
\usepackage{caption}
\usepackage{algpseudocode} 
\usepackage{xcolor}
\usepackage{booktabs}
\usepackage{multirow}
\usepackage{wrapfig}

\usepackage{colortbl} 
\usepackage[scaled]{inconsolata}
\usepackage{multirow}
\usepackage{array}
\usepackage[table]{xcolor}
\usepackage{subcaption}
\usepackage{graphicx}
\usepackage{capt-of}
\usepackage{caption}
\usepackage{pifont}
\usepackage{bbding}
\usepackage{fontawesome5}
\usepackage{tikz}
\usepackage{eso-pic}

\usepackage[normalem]{ulem}
\newcommand{\cmark}{\textcolor{green!60!black}{\CheckmarkBold}}
\newcommand{\xmark}{\textcolor{red!75!black}{\XSolidBrush}}
\usepackage{makecell}

\usepackage{titlesec}
\definecolor{headerblue}{RGB}{220,235,250}
\definecolor{deepblue}{RGB}{0, 45, 114}
\usepackage{fontawesome5}
\usepackage{hyperref}
\usepackage{xcolor}
\usepackage{graphicx}

\title{Mega-ASR: Towards \textit{In-the-wild}$^{2}$ Speech Recognition  via Scaling Up Real-world Acoustic Simulation}

\vspace{-2mm}
\author{%
  Zhifei Xie\textsuperscript{1*},\quad Kaiyu Pang\textsuperscript{3*},\quad Haobin Zhang\textsuperscript{2*},\quad Deheng Ye\textsuperscript{1$\dagger$},\quad
  Xiaobin Hu\textsuperscript{2$\dagger$} \\
  \textbf{Shuicheng Yan}\textsuperscript{2$\dagger$},\quad \textbf{Chunyan Miao}\textsuperscript{1$\dagger$} \\
  \textsuperscript{1}NTU \quad \textsuperscript{2}NUS \quad \textsuperscript{3}Shanghai AI Lab \\
  \faEnvelope\space \href{mailto:Zhifei001@e.ntu.edu.sg}{\texttt{Zhifei001@e.ntu.edu.sg}}%
}

\begin{document}

\AddToShipoutPictureFG*{%
  \AtPageUpperLeft{%
    \hspace{1.5in}%
    \raisebox{-1.1in}{%
      \includegraphics[width=1.5cm]{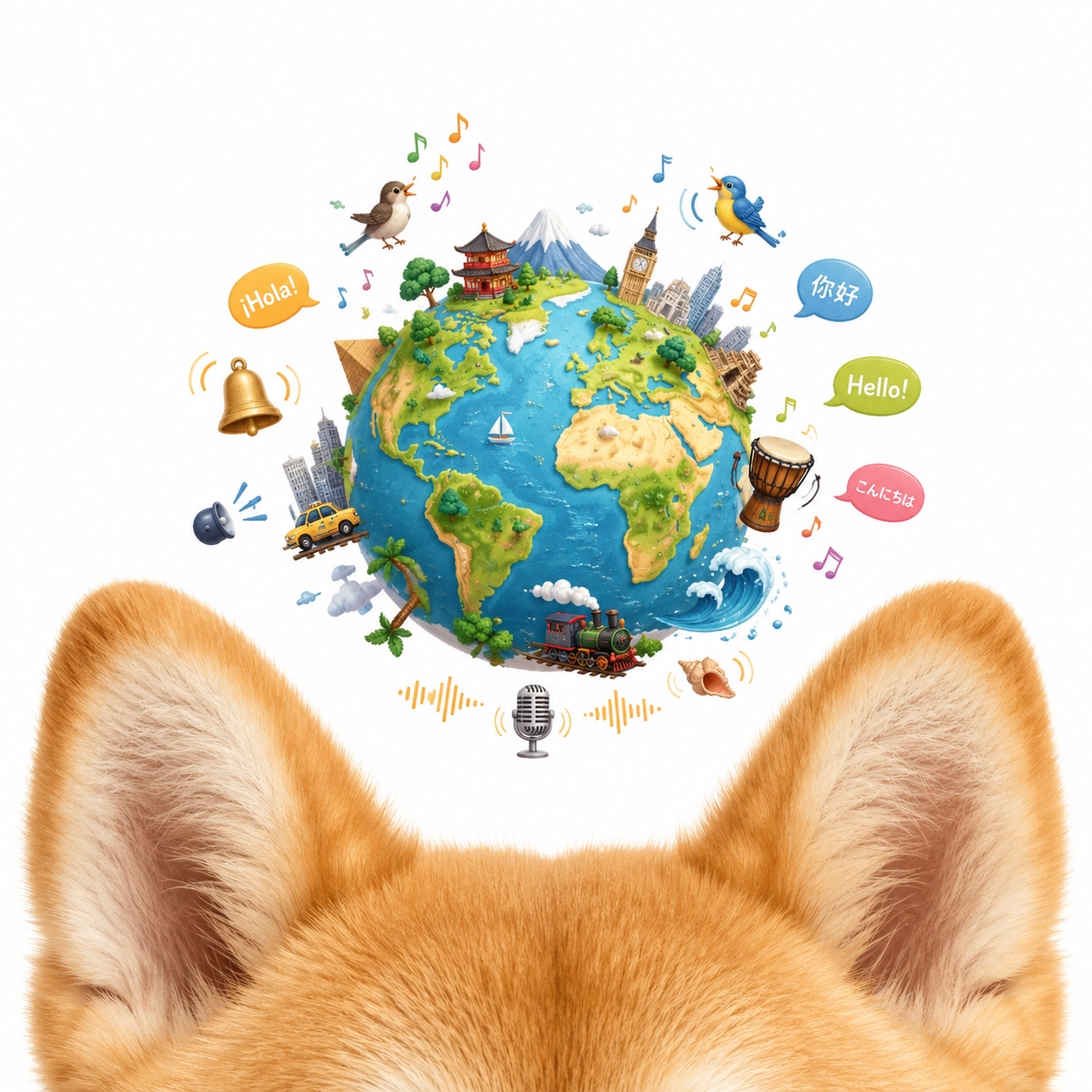}%
    }%
  }%
}

\maketitle
\vspace{-7mm}

\newcommand{\logoblog}{\raisebox{-0.2ex}{\includegraphics[height=1em]{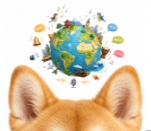}}}
\newcommand{\logohf}{\raisebox{-0.2ex}{\includegraphics[height=1em]{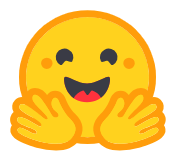}}}
\newcommand{\logogh}{\raisebox{-0.2ex}{\includegraphics[height=1em]{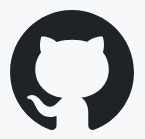}}}

\vspace{-2mm}
\begin{center}
\begin{minipage}{0.7\textwidth}
\logoblog~\textbf{Project page:} \href{https://xzf-thu.github.io/Mega-ASR/}{\texttt{\textcolor{deepblue}{https://xzf-thu.github.io/Mega-ASR/}}} \\
\logohf~\textbf{Data:} \href{https://huggingface.co/datasets/zhifeixie/Voices-in-the-Wild-2M}{\texttt{\textcolor{deepblue}{huggingface.co/datasets/zhifeixie/Voices-in-the-Wild-2M}}} \\
\logogh~\textbf{Bench:} \href{https://github.com/xzf-thu/Voices-in-the-Wild-Bench}{\texttt{\textcolor{deepblue}{github.com/xzf-thu/Voices-in-the-Wild-Bench}}}
\end{minipage}
\end{center}

\vspace{1mm}
\begin{abstract}
Despite rapid advances in automatic speech recognition (ASR) and large audio-language models, robust recognition in real-world environments remains limited by an ``acoustic robustness bottleneck'': models often lose acoustic grounding and produce omissions or hallucinations under severe, compositional distortions. We propose \textbf{MEGA-ASR}, a unified ASR-in-the-wild framework that combines scalable compound-data construction with progressive acoustic-to-semantic optimization. We introduce \textbf{VOICES-IN-THE-WILD-2M}, covering \textbf{7} classic acoustic phenomena and \textbf{54} physically plausible compound scenarios, and train MEGA-ASR with \textbf{Acoustic-to-Semantic Progressive Supervised Fine-Tuning} and \textbf{Dual-Granularity WER-Gated Policy Optimization}. Extensive experiments demonstrate that MEGA-ASR achieves significant advantages over prior state-of-the-art systems on adverse-condition ASR benchmarks (45.69\% vs. 54.01\% on VOiCES R4-B-F, and 21.49\% vs. 29.34\% on NOIZEUS Sta-0). On complex compositional acoustic scenarios, MEGA-ASR further delivers \textbf{over 30\%}relative WER reduction against strong open- and closed-source baselines, establishing a scalable paradigm for robust ASR in-the-wild.
\end{abstract}

\vspace{-2mm}
\begin{figure}[!h]
    \centering
    \makebox[\textwidth][c]{%
        \includegraphics[width=1.1\textwidth]{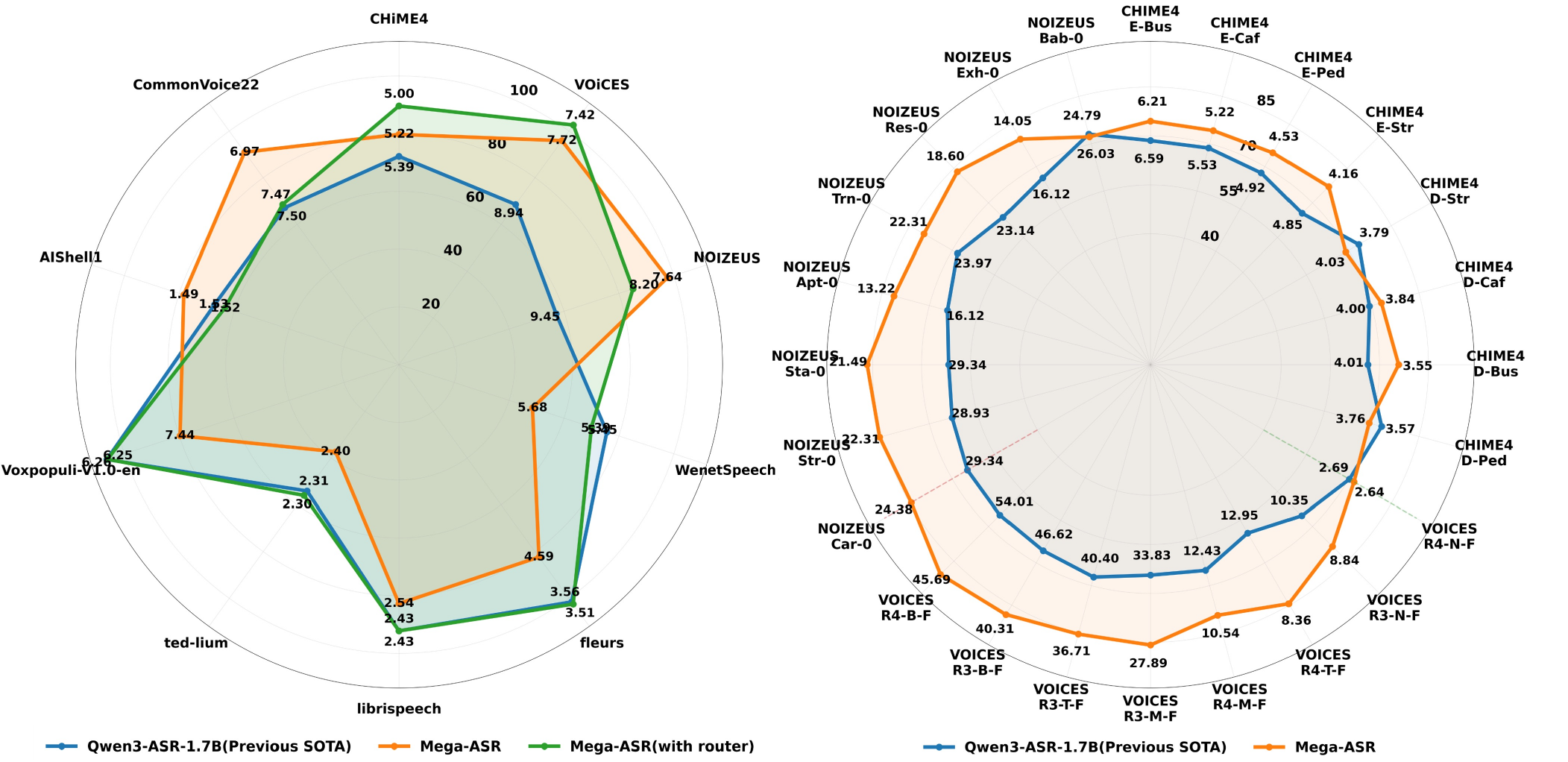}%
    }
    \caption{
        Radar comparison of Qwen3-ASR-1.7B and Mega-ASR across selected ASR evaluation subsets, covering both clean and robustness benchmarks.
    }
    \label{fig:radar_comparison}
\end{figure}
\vspace{-1em}








\section{Introduction}
\vspace{-1mm}
Automatic speech recognition (ASR) is one of the most fundamental tasks in the speech domain, and has evolved rapidly in recent years. State-of-the-art ASR models~\citep{qwen3-asr,fireredasr2s,paraformer} achieve excellent accuracy on widely used benchmarks~\citep{librispeech}, with word error rates approaching 1\%. Beyond this, large audio-language models (LALMs)~\citep{qwen3-omni,Kimi-Audio} scale to billion-parameter architectures that integrate pretrained linguistic knowledge and even support reasoning-based error correction~\citep{reasoningforasr}, improving contextual consistency and achieving human-level performance on canonical benchmarks.
\vspace{-0.5mm}

However, performance drops sharply under real-world acoustic conditions: WER typically rises to 10\%--30\%, and in harder cases can be as high as 
70\%, often accompanied by \underline{\textit{dropped utterances or}} \underline{\textit{severe hallucination}}. Recent work on \emph{ASR-in-the-wild}~\citep{yan2025improving,asrinthewild} seeks to bridge this gap through improved data and post-training strategies. Nevertheless, three limitations persist.
\textbf{(D1) Limited scenario coverage.} Prior work typically targets one or two isolated conditions (e.g., noise or far-field), requiring different specialized models for different environments. 
\textbf{(D2) Lack of compositional robustness.} Robustness factors are studied independently, while real-world conditions are inherently compositional (e.g., simultaneous reverberation, echo, and frequency dropout), and large-scale data for such mixtures remains scarce. 
\textbf{(D3) Mismatch between training data and real-world conditions.} The data that existing models are trained on emphasize relatively mild WER ranges ($4\%$--$10\%$), which do not reflect challenging settings where WER exceeds 30\% and demands stronger semantic reasoning over degraded signals. These gaps motivate a shift toward \emph{ASR-in-the-wild$^2$}, pushing ASR models to handle acoustic conditions that are not just singly complex, and to recognize speech under much harder settings.
\vspace{-0.5mm}

In this work, we propose \textbf{\textsc{Mega-ASR}}, a framework specifically designed to strengthen ASR capability under \textit{in-the-wild} complex acoustic environments. \textsc{Mega-ASR} is able to \textbf{(1)} achieve state-of-the-art accuracy on individual environmental conditions within a single model, \textbf{(2)} deliver superior performance on real-world recordings exhibiting compound environmental effects, and \textbf{(3)} recover semantic information under highly challenging conditions, which requires a dataset that is both close to the real-world distribution and scalable. To this end, we introduce \textbf{\textsc{ \textsc{Voices-in-the-wild-2M} }}, a large-scale ASR dataset comprising \textbf{7} canonical meta-scenarios and \textbf{54} newly constructed compound scenarios, generated by a spectral-manipulation-based simulation method. We first  \underline{\textit{(i) simulate 7}} \underline{\textit{atomic acoustic effects}} in isolation as the foundation, then \underline{\textit{(ii) scale to 54 compound scenarios}} with an agentic check that verifies physical plausibility (e.g., a church corresponds to far-field plus echo).  To obtain data that is both challenging and suitable for training, we \underline{\textit{(iii) calibrate the difficulty}} \underline{\textit{distribution}} through controlled experiments, and finally \underline{\textit{(iv) filter out samples with WER above 70\%}} to ensure training stability. We then develop \textbf{Acoustic-to-Semantic Progressive Supervised Fine-Tuning (A2S-SFT)}, addressing two coupled bottlenecks at medium-to-high WER: extracting semantic information from acoustic signals under heavy perturbation, and recovering the intended semantics. Through this progressive capability building, we obtainx \textbf{\textsc{Mega-ASR-Base}}, whose foundational capabilities for the reward signal that subsequent reinforcement learning depends on.

\vspace{-0.5mm}

Finally, during RL training, recognition errors at medium difficulty are mostly word-level mistakes, but once WER exceeds 30\%, the dominant failure mode changes sharply into severely incorrect semantics, hallucinated guesses, and large portions of dropped sentences. As a result, WER-based rewards cannot provide an effective learning signal in this situation. We therefore propose \textbf{Dual-Granularity WER-Gated Policy Optimization (DG-WGPO)}, a dynamic reward scheme with two parts. We also adopt a classic \textbf{static rule-based reward} consisting of WER and a repetition penalty as the basic learning signal. As the core of DG-WGPO, we introduce a \textbf{Dual-Granularity Dynamic Reward} designed specifically for ASR under complex acoustic environments, which combines a \underline{\textit{token-level refinement reward for local information recovery}} and a \underline{\textit{sentence-level reconstruction reward for overall semantic preservation on hard samples}}, with a \underline{\textit{WER-gated mirrored fusion strategy}} that dynamically allocates the weights between them. Extensive experiments show that MEGA-ASR substantially outperforms prior state-of-the-art systems on adverse-condition and compositional real-world benchmarks.

\vspace{-2mm}
\section{Related Work}
\vspace{-2mm}
\paragraph{ASR Foundation Models and Robust Speech Recognition.}
Recent ASR foundation models, spanning encoder-decoder systems, large-scale self-supervised models, and audio-language models, have achieved strong results on standard benchmarks~\citep{whisper, fireredasr2s, qwen3-asr, funasr, qwen2.5omni, qwen3-omni, Kimi-Audio, stepaudio2}. However, strong performance under clean or mildly noisy conditions does not imply robustness in deployment, where speech is often corrupted by simultaneous degradations such as noise, far-field propagation, reverberation, obstructed, device distortion, and transmission dropout. Existing robust ASR studies typically address only one or two such factors, leaving severe and compositional conditions underexplored.
\vspace{-4mm}
\paragraph{Datasets and Simulation for In-the-wild ASR.}
A long line of robust ASR benchmarks studies recognition under adverse conditions, including additive noise, distant microphones, reverberation, replayed speech, and device effects~\citep{Noizeus, CHIME4, VOICES, DAPS, ted-lium, commonvoice, voxpopuli}, but most emphasize isolated factors or mild degradation regimes. In practice, environments such as classrooms, corridors, or vehicles routinely combine background noise, far-field attenuation, echo, occlusion, and device-induced distortion. Augmentation methods like noise mixing, RIR convolution, spectral masking, clipping, and codec simulation partially address this~\citep{MUSAN, DNS, ko2015audio, ko2017audio, parada2022pmctpatchedmulticonditiontraining}, but typically serve as local training perturbations rather than a systematic model of real acoustic worlds.

\vspace{-2mm}
\section{ \textsc{Voices-in-the-wild-2M} }
\vspace{-3mm}
\subsection{Overview}
\vspace{-2mm}

\begin{figure}
    \centering
    \includegraphics[width=0.96\linewidth]{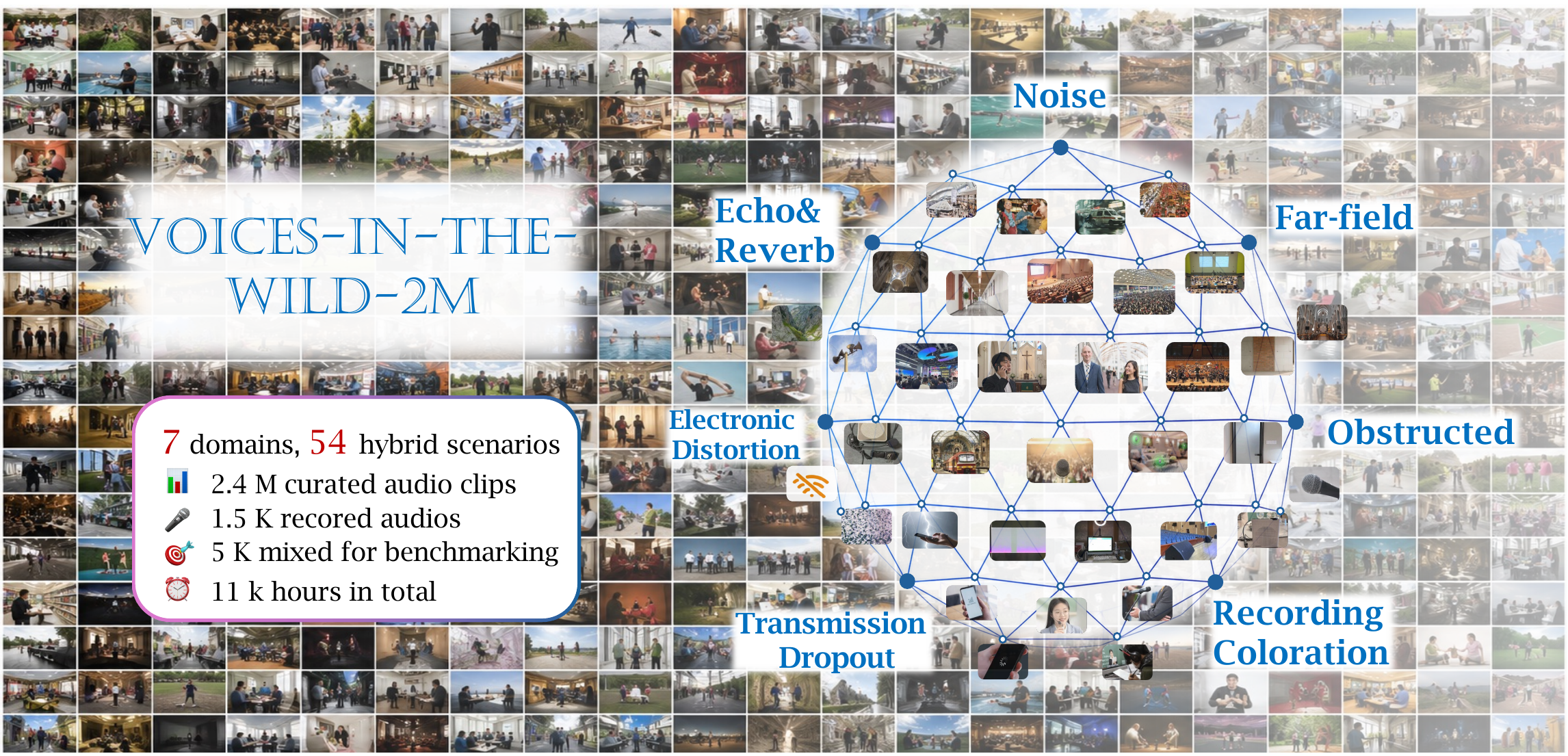}
    \caption{ \textsc{Voices-in-the-wild-2M}  enables environmentally robust ASR by expanding 7 meta-scenarios into 54 hybrid scenarios, covering diverse real-world acoustic degradations at scale.}
    \vspace{-3mm}
    \label{fig:dataset-overview}
\end{figure}

Existing datasets for robust ASR mostly cover only a narrow set of isolated acoustic conditions, with mild WER typically between\textbf{ 4\%--10\%} as shown in Table~\ref{table:relatedworks}, whereas real-world environments mix multiple environmental effects (e.g., far-field with echo\&reverb in a church interior) and routinely push WER beyond 30\%. To facilitate research in this regime, we introduce \textbf{ \textsc{Voices-in-the-wild-2M} }, a large-scale dataset built through spectrogram-level code-based simulation, the design choice that makes its scale tractable. To faithfully simulate the complex acoustic conditions encountered in-the-wild, we first identify, as shown in Figure~\ref{fig:dataset-overview}, seven classic in-the-field acoustic effects $\left\{\textit{noise},\ \textit{far-field},\ \textit{obstructed},\ \textit{echo\&reverb},\ \textit{recording},\ \textit{electronic distortion},\ \textit{transmission dropout}\right\}$, which we term \underline{\textit{atomic acoustic effects}}. Each atomic effect is implemented as a dedicated spectral processing pipeline and iteratively calibrated against real recordings, with parameters re-tuned and validated via SFT on Qwen3-ASR until the simulator attains best fit on real data. The atomic phenomena are then composed into \textbf{54} agent-validated configurations, yielding \textbf{2.4M} synthesized clips whose effectiveness on real-world data is empirically verified after mixed-condition training.  \textsc{Voices-in-the-wild-2M}  is also substantially more challenging, thereby promoting robustness in complex real-world environments: even the state-of-the-art Qwen3-ASR~\citep{qwen3-asr} attains a high average WER of \textbf{35\%} on this benchmark.

\begin{table}[t]
\centering
\footnotesize
\setlength{\tabcolsep}{3.5pt}
\renewcommand{\arraystretch}{1.05}
\caption{Coverage comparison of acoustic degradation scenarios across datasets.}
\begin{tabular}{lccccccccccc}
\toprule
 & \multicolumn{2}{c}{\textbf{source}} & \multicolumn{7}{c}{\textbf{Acoustic Phenomena}} &  &  \\
\cmidrule(lr){2-3} \cmidrule(lr){4-10}
\textbf{Dataset} & \textbf{real.} & \textbf{sim.} & \textbf{Noise} & \textbf{Far} & \textbf{Barr.} & \textbf{E\&R} & \textbf{Record} & \textbf{Distort} & \textbf{Drop} & \textbf{Scale} & \textbf{WER} \\
\midrule
NOIZEUS \citep{Noizeus}   & \xmark & \cmark & \cmark & \xmark & \xmark & \xmark & \xmark & \xmark & \xmark & 1K   & 9.45 \\
TED-LIUM \citep{ted-lium}  & \cmark & \xmark & \xmark & \cmark & \xmark & \xmark & \xmark & \xmark & \xmark & 59K  & 2.31 \\
CHiME-4 \citep{CHIME4}  & \cmark & \cmark & \cmark & \xmark & \xmark & \xmark & \xmark & \xmark & \xmark & 15K  & 5.39 \\
VOiCES \citep{CHIME4}& \cmark & \xmark & \cmark & \cmark & \cmark & \cmark & \cmark & \xmark & \xmark & 1M   & 8.94 \\
BERSt \citep{BERSt}     & \cmark & \xmark & \xmark & \cmark & \cmark & \xmark & \xmark & \xmark & \xmark & 4.5K & 22.41 \\
DAPS \citep{DAPS}     & \cmark & \xmark & \xmark & \xmark & \xmark & \xmark & \xmark & \cmark & \xmark & 2K   & 6.24 \\
\textbf{\textsc{\textsc{Voices-in-the-wild-2M} } }    & \cmark & \cmark & \cmark & \cmark & \cmark & \cmark & \cmark & \cmark & \cmark & 2M   & 18.42 \\
\bottomrule
\end{tabular}
\vspace{-2mm}
\label{table:relatedworks}
\end{table}

\vspace{-2mm}
\subsection{Realistic Simulation of Compound Acoustic Environments}
\begin{figure}[!h]
    \centering
    \includegraphics[width=\textwidth]{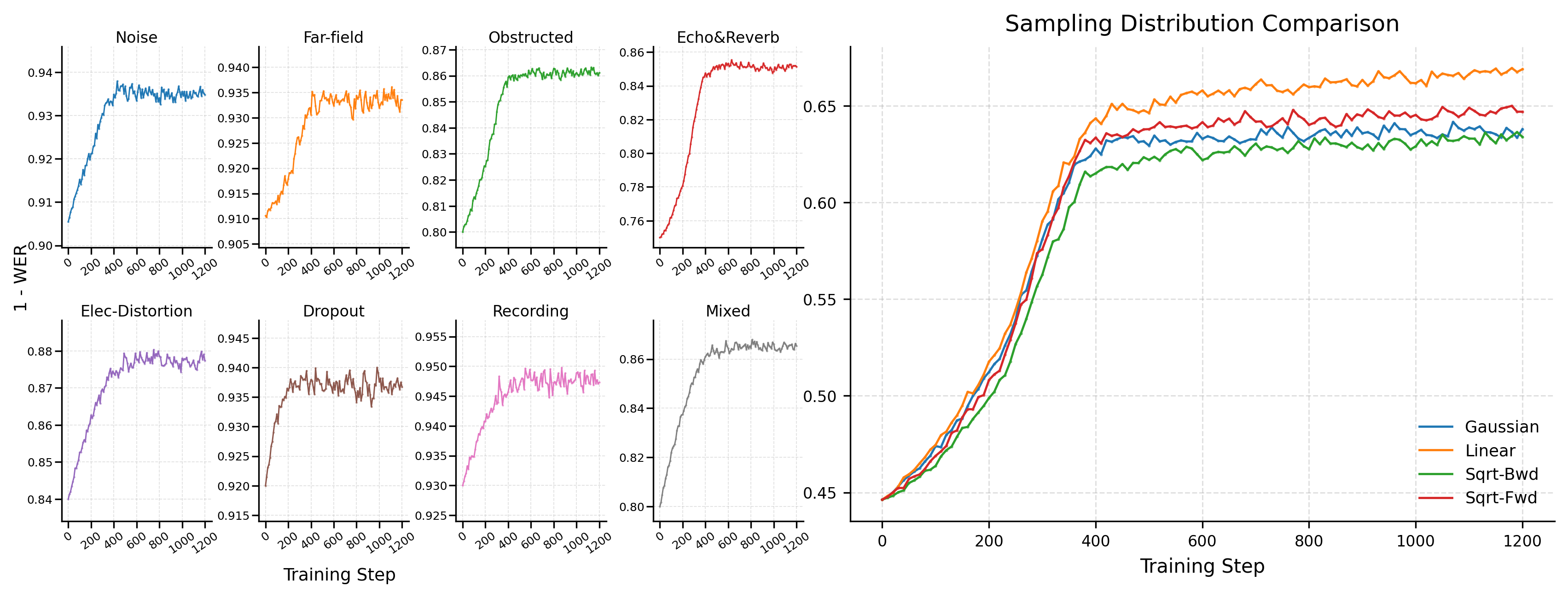}
    \vspace{-7mm}
    \caption{Left: SFT accuracy curves on real samples after careful tuning, shown for individual and mixed atomic effects. Right: comparison of difficulty sampling distributions on Noizeus 0dB.}
    \vspace{-2mm}
    \label{fig:robustness_sampling_combined}
\end{figure}

\vspace{-1mm}
In principle, two routes exist for building such a dataset: 
\textit{(Option 1) curating existing materials such as online videos}, which 
we found costly and fundamentally unscalable, and \textit{(Option 2) synthesizing from clean 
speech clips}. We adopt the latter for its flexibility and, more importantly, 
its scalability. The pipeline proceeds as follows.
\textit{(i) Atomic acoustic effect simulation.} As the foundation of the pipeline, we simulate each of the seven phenomena directly on the spectrogram via filtering, convolution, and related signal-level transformations, with parameters iteratively tuned to best fit real-world recordings. We further incorporate a broad collection of real-world material spanning comprehensive background and speech sources: noise from MUSAN~\citep{MUSAN}, DNS Challenge~\citep{DNS}, ESC-50~\citep{ESC-50}, and UrbanSound8K~\citep{URBANSOUND} (\textasciitilde42K clips, 129 hours), and clean speech from LibriSpeech~\citep{librispeech}, Common Voice~\citep{VOICES}, WenetSpeech~\citep{wenetspeech}, and AISHELL-1~\citep{aishell}.
\textit{(ii) Reality-grounded composition.} Since real environments rarely exhibit a single isolated effect, we scale from atomic effects to compound scenarios by composing 2 to 5 atomic effects, retaining only physically plausible combinations (e.g., far-field with ambient noise in 
a church interior) and yielding the 54 compound configurations above. 
\textit{(iii) Controllable-difficulty synthesis.} To obtain data that is both challenging and suitable for training, we calibrate the difficulty distribution by exposing a unified severity 
parameter $k\in[0,1]$ for every effect and generating 50K probe samples under 
four candidate distributions over $k$ (Sqrt-Forward, Sqrt-Backward, 
Gaussian-Mid, Linear); as shown in Figure~\ref{fig:robustness_sampling_combined}, the \underline{\textit{Linear 
distribution is adopted as the severity profile of the dataset}}. 
\textit{(iv) Learnability fi-}

\vspace{-4mm}

\vspace{-1mm}
 \textit{tering.} To ensure training stability, we discard samples with WER above 70\%, 
which we observe to destabilize training otherwise. Full pipeline details and examples are provided in the appendix~\ref{app:data_construction}.

\vspace{-3mm}
\subsection{Voices-in-the-wild-Bench: A Real-Recording Evaluation Benchmark}
\vspace{-2mm}
We further release Voices-in-the-wild-Bench, a 5{,}000-clip English/Mandarin evaluation set covering the same seven atomic phenomena as \textsc{Voices-in-the-wild-2M}, comprising 3{,}500 synthetic clips and 1{,}500 real-world recordings collected from internet sources and 16 human participants.

\vspace{-2mm}
\section{Mega-ASR}
\begin{figure}
    \centering
    \includegraphics[width=0.95\linewidth]{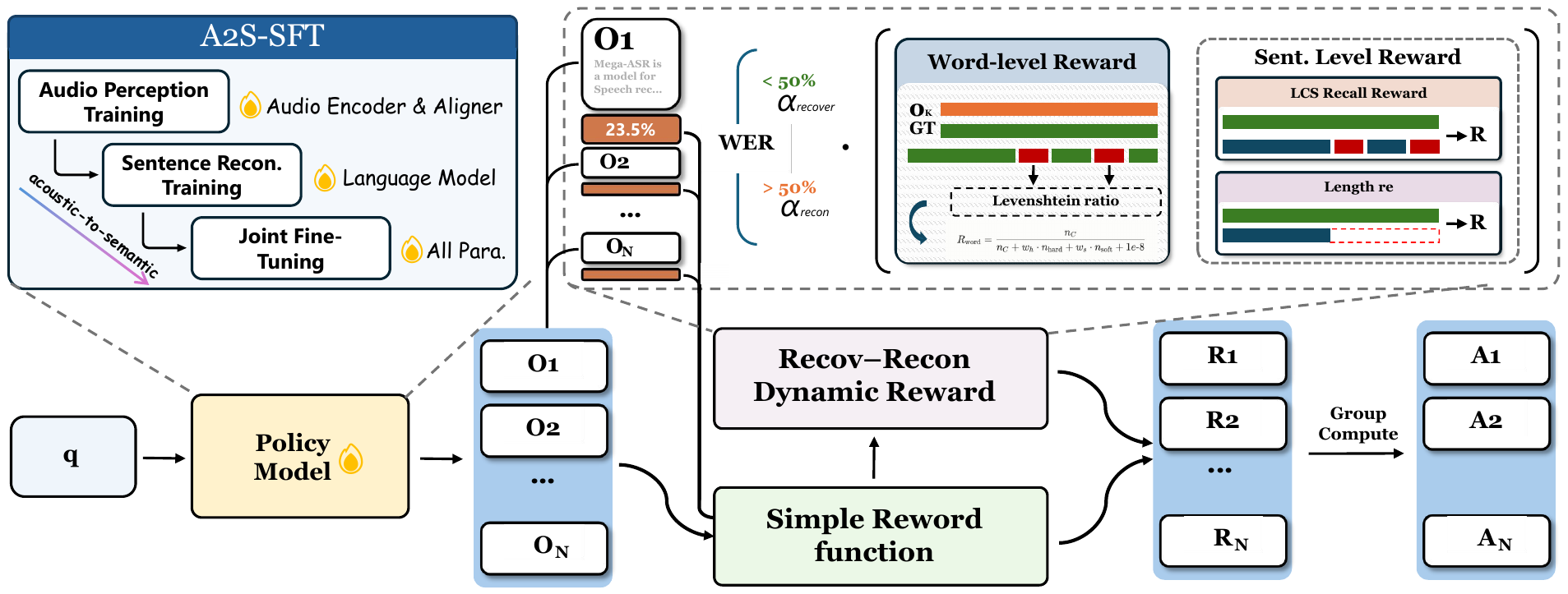}
    \caption{Overview of the proposed DG-WGPO framework. Starting from A2S-SFT initialization, the policy model generates multiple hypotheses scored by a dynamic reward with gated fusion.}
    \vspace{-2mm}
    \label{fig:modeltraining}
\end{figure}

\vspace{-2mm}
We propose a framework, as shown in figure~\ref{fig:modeltraining} for robust speech recognition under complex acoustic conditions. We first develop \textbf{Mega-ASR-Base} on top of Qwen3-ASR~\citep{qwen3-asr} via \textbf{Acoustic-to-Semantic Progressive Supervised Fine-Tuning}, instilling perceptual robustness and semantic recovery.We then apply \textbf{Dual-Granularity WER-Gated Policy Optimization} that supplies token- and sentence-level rewards, dynamically modulating their granularity to mitigate WER reward failure.

\vspace{-2mm}
\subsection{Acoustic-to-Semantic Progressive Supervised Fine-Tuning}
\vspace{-1mm}
We observe that existing ASR models struggle to maintain reliable acoustic understanding in the medium and high WER regimes, often producing \underline{\textit{empty outputs, severe hallucinations, or off-audio}} \underline{\textit{transcriptions.}} The failure stems from two coupled bottlenecks: \textit{(i)} extracting reliable acoustic evidence from corrupted waveforms, which the encoder-aligner stack alone cannot guarantee, and \textit{(ii)} leveraging the LLM's semantic prior to reconstruct the intended transcription when that evidence is only partially reliable. A2S-SFT addresses them in three phases: \textit{(i)} a WER-graded curriculum on the encoder and aligner, successively expanding from $\text{WER}{<}30\%$ to $\text{WER}{<}50\%$ and finally to $\text{WER}{<}70\%$, to build acoustic perception incrementally; \textit{(ii)} LLM fine-tuning on full $\text{WER}{<}70\%$ samples to activate semantic recovery under unreliable acoustic evidence; and \textit{(iii)} joint fine-tuning of encoder, aligner, and LLM for end-to-end alignment.
\vspace{-2mm}
\subsection{Dual-Granularity WER-Gated Policy Optimization}
\vspace{-2mm}
Building on \textbf{Mega-ASR-Base}, we apply DAPO~\citep{dapo} to sharpen the policy.\textit{ We observe during training that errors when $\text{WER}{<=}30\%$ are predominantly word-level confusions, whereas beyond this threshold they shift abruptly into sentence-level failures such as hallucinations and omissions.} The standard WER reward, however, conflates these two regimes and further saturates under heavy degradation, collapsing intra-group dispersion precisely where the policy needs it most. We therefore propose \textbf{Dual-Granularity WER-Gated Policy Optimization (DG-WGPO)}, which retains a classic \textbf{static rule-based reward} (WER plus a repetition penalty) as the basic learning signal, and introduces a \textbf{Dual-Granularity Dynamic Reward} as its core, applying WER-gated fine- and coarse-grained rewards aligned with the two error regimes.
\vspace{-2mm}
\subsubsection{Static Rule-Based Rewards}
\vspace{-1mm}
The static rewards provide a stable, sample-independent anchor that ties the policy directly to the evaluation metric while filtering out degenerate rollouts.
\paragraph{WER reward.} The WER reward serves as a direct anchor to the evaluation metric:
\begin{equation}
R_\text{wer}(H, R) = 1 - \text{WER}(H, R).
\end{equation}
\paragraph{Anti-repetition reward.} Rollouts occasionally collapse into repeated short n-grams, inflating token coverage with hallucinated content. We apply a multiplicative hard gate that zeros out such rollouts:
\begin{equation}
R_\text{rep}(H) = 
\begin{cases}
0, & \text{if } H \text{ contains repeated $n$-grams beyond threshold}, \\
1, & \text{otherwise}.
\end{cases}
\end{equation}
We aggregate the two into a single static signal that gates transcription accuracy on non-degenerate rollouts:
\begin{equation}
R_\text{static} = R_\text{rep} \cdot R_\text{wer}.
\end{equation}

\vspace{-1mm}
\subsubsection{Dual-Granularity Dynamic Reward}
\vspace{-1mm}
At the core of DG-WGPO, the Dual-Granularity Dynamic Reward is designed specifically for ASR under complex acoustic environments. It combines a token-level refinement reward for local information recovery and a sentence-level reconstruction reward for overall semantic preservation on hard samples, with a WER-gated mirrored fusion strategy that dynamically allocates the weights between them.

\vspace{-1mm}
\paragraph{Token-level refinement reward.} Targeting failure mode \textit{(i)}, we partition substitution errors by character-level edit similarity. Given a hypothesis token $h$ and reference token $r$,
\begin{equation}
\text{sim}(h,r) = 1 - \frac{\text{edit}(h,r)}{\max(|h|,|r|)} \in [0,1],
\end{equation}
and we classify a substitution as \textit{soft} if $\text{sim}(h,r) \geq 0.5$ (the midpoint of the similarity range) and \textit{hard} otherwise. Insertions and deletions are uniformly treated as hard, since both signal hallucination rather than acoustic confusion. The refinement reward discounts the two error types separately:
\begin{equation}
R_\text{fine} = \frac{n_C}{n_C + n_\text{hard} + \alpha_s\, n_\text{soft} + \epsilon},
\end{equation}
where $n_C$, $n_\text{hard}$, $n_\text{soft}$ are the counts of correct tokens, hard errors, and soft errors respectively, $\alpha_s \in (0,1)$ is the soft-error discount, and $\epsilon = 10^{-8}$ ensures numerical stability.

\vspace{-1mm}
\paragraph{Sentence-level reconstruction reward.} Targeting failure mode \textit{(ii)}, we score the hypothesis by backbone preservation rather than token-level agreement:
\begin{equation}
R_\text{struc} = \frac{1}{2} \cdot \frac{\text{LCS}(H,R)}{|R|} + \frac{1}{2} \cdot \max\!\left(0,\, 1 - \frac{\big||H|-|R|\big|}{|R|}\right),
\end{equation}
where the LCS term rewards backbone agreement under local reordering and the length term penalizes truncation and runaway generation. The two terms are equally weighted as both contribute to structural integrity.

\vspace{-1mm}
\paragraph{WER-gated dynamic fusion.} The relative usefulness of the two granularities flips at the refinement-reconstruction boundary, so we fuse them with a WER-gated mirrored weighting that always assigns the dominant weight to the regime-appropriate granularity:
\begin{equation}
R_\text{dynamic} =
\begin{cases}
0.75\, R_\text{fine} + 0.25\, R_\text{struc}, & \text{WER}(H,R) < \tau, \\[2pt]
0.25\, R_\text{fine} + 0.75\, R_\text{struc}, & \text{WER}(H,R) \geq \tau.
\end{cases}
\end{equation}

\paragraph{Final objective.} The full reward combines the rule-based anchor with the dynamic signal:
\begin{equation}
R = (1 - \alpha_\text{dyn})\, R_\text{simple} + \alpha_\text{dyn}\, R_\text{dynamic}.
\end{equation}
We set the three hyperparameters as $\tau = 0.3$, $\alpha_s = 0.4$, and $\alpha_\text{dyn} = 0.6$.

\vspace{-2mm}
\subsection{Environment-Aware Routing for Plug-and-Play Inference}
\vspace{-2mm}

\noindent
\begin{minipage}[t]{0.55\linewidth}
Training \textsc{Mega-ASR} on heavily degraded audio sharpens its noise robustness but partially erodes
complementary capabilities such as clean-speech recognition, hotword recognition, and streaming ASR.
To preserve both, we route each utterance to the appropriate model at inference time.
Specifically, as illustrated in figure~\ref{fig:inference-routing} we fine-tune a lightweight binary classifier with LoRA on a mixture of clean speech and
\textit{Voices-in-the-Wild} samples, predicting whether an input requires Mega-ASR's noise-robust
weights or the original backbone. This routing keeps \textsc{Mega-ASR} as a plug-and-play module that activates
only when the acoustic environment demands it, leaving clean-domain performance untouched.
\end{minipage}
\hfill
\begin{minipage}[t]{0.41\linewidth}
    \centering
    \vspace{0pt}
    \includegraphics[width=\linewidth]{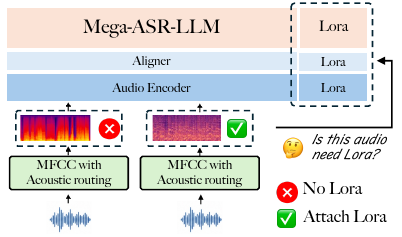}
    \captionof{figure}{Environment-aware routing for plug-and-play inference.}
    \label{fig:inference-routing}
\end{minipage}

\vspace{1mm}

\begin{table*}[t]
\centering
\footnotesize
\renewcommand{\arraystretch}{1.0}

\caption{Performance comparison on noisy and robust ASR benchmarks.}
\vspace{-2mm}
\label{tab:main_noise}

\setlength{\tabcolsep}{2.73pt}
\begin{tabular}{lcccccccccccccc}
\toprule
\multirow{2}{*}{\textbf{Model}} &
\multicolumn{3}{c}{\textbf{CHiME-4}} &
\multicolumn{5}{c}{\textbf{VOiCES}} &
\multicolumn{5}{c}{\textbf{NOIZEUS}} &
\multirow{2}{*}{\textbf{Avg.}} \\
\cmidrule(lr){2-4}
\cmidrule(lr){5-9}
\cmidrule(lr){10-14}
&
\textbf{Real} &
\textbf{Sim} &
\textbf{Avg.} &
\textbf{rm1} &
\textbf{rm2} &
\textbf{rm3} &
\textbf{rm4} &
\textbf{Avg.} &
\textbf{0dB} &
\textbf{5dB} &
\textbf{10dB} &
\textbf{15dB} &
\textbf{Avg.} &
\\
\midrule
\rowcolor{headerblue}
\multicolumn{15}{l}{\textit{Closed-source models}} \\
Gemini3-Flash &
6.58 &
5.67 &
6.125 &
3.10 &
4.27 &
25.99 &
21.86 &
13.81 &
55.78 &
24.48 &
18.49 &
8.52 &
26.82 &
15.59 \\
Doubao-LLM ASR &
9.95 &
11.62 &
10.79 &
4.86 &
6.99 &
17.23 &
7.85 &
9.23 &
25.78 &
9.51 &
4.96 &
2.87 &
10.78 &
10.27 \\
GPT-4o-trans. &
5.36 &
7.57 &
6.47 &
10.97 &
12.56 &
46.68 &
29.38 &
22.65 &
62.40 &
20.56 &
6.15 &
2.64 &
22.94 &
17.35 \\
\rowcolor{headerblue}
\multicolumn{15}{l}{\textit{Open-source models}} \\
Voxtral-Mini &
6.01 &
9.04 &
7.53 &
3.50 &
3.51 &
27.54 &
16.45 &
12.75 &
41.06 &
15.80 &
4.85 &
2.94 &
16.16 &
12.15 \\
Kimi-Audio &
5.66 &
7.46 &
6.56 &
\underline{2.10} &
\underline{2.23} &
26.95 &
15.13 &
11.60 &
38.33 &
11.36 &
4.34 &
2.27 &
14.08 &
10.74 \\
Whisper-L-v3 &
5.65 &
8.39 &
7.02 &
2.85 &
2.97 &
25.68 &
15.65 &
11.79 &
34.71 &
12.55 &
3.93 &
2.17 &
13.34 &
10.72 \\
Canary-1B-v2 &
7.19 &
9.73 &
8.46 &
3.14 &
3.00 &
24.88 &
15.56 &
11.65 &
38.53 &
12.76 &
6.56 &
3.77 &
15.41 &
11.84 \\
Parakeet-v3 &
6.61 &
8.82 &
7.72 &
3.23 &
3.27 &
19.77 &
13.84 &
10.03 &
38.95 &
14.67 &
5.99 &
3.15 &
15.69 &
11.15 \\
Qwen2.5-Omni &
6.62 &
8.13 &
7.37 &
4.15 &
4.03 &
44.76 &
22.53 &
18.87 &
54.91 &
17.72 &
3.20 &
\textbf{0.88} &
19.18 &
15.14 \\
Step-Audio-2-mini &
5.35 &
7.06 &
6.20 &
\textbf{1.81} &
\textbf{1.98} &
23.25 &
15.19 &
10.56 &
32.02 &
8.94 &
3.72 &
2.27 &
11.74 &
9.50 \\
Qwen3-ASR &
4.66 &
6.11 &
5.39 &
2.52 &
2.62 &
19.18 &
11.44 &
8.94 &
\underline{23.97} &
8.47 &
3.41 &
1.96 &
9.45 &
7.93 \\
\rowcolor{headerblue}
\multicolumn{15}{l}{\textit{Our model}} \\
Mega-ASR &
\underline{4.41} &
\underline{6.04} &
\underline{5.23} &
2.36 &
2.43 &
\textbf{15.13} &
\underline{9.46} &
\textbf{7.35} &
\textbf{19.80} &
\textbf{6.61} &
\textbf{2.79} &
\textbf{0.88} &
\textbf{7.52} &
\textbf{6.70} \\
Mega-ASR w/ router &
\textbf{4.38} &
\textbf{5.62} &
\textbf{5.00} &
2.42 &
2.49 &
\underline{15.32} &
\textbf{9.26} &
\underline{7.37} &
\textbf{19.80} &
\underline{6.97} &
\underline{3.05} &
\underline{1.76} &
\underline{7.90} &
\underline{6.76} \\
\bottomrule
\end{tabular}

\vspace{2mm}

\caption{Performance comparison on standard ASR benchmarks. For LibriSpeech, each entry is reported as clean/other. Underline indicates the best performance among open-source models.}
\vspace{-2mm}
\label{tab:main_standard}
\setlength{\tabcolsep}{3.9pt}
\begin{tabular}{lcccccccccc}
\toprule
\textbf{Model} &
\multicolumn{2}{c}{\textbf{LibriSp.}} &
\multicolumn{2}{c}{\textbf{Comm.Voice}} &
\multicolumn{2}{c}{\textbf{Fleurs}} &
\textbf{AISHELL-1} &
\multicolumn{2}{c}{\textbf{WenetSp.}} &
\textbf{VoxPop.} \\
\cmidrule(lr){2-3}
\cmidrule(lr){4-5}
\cmidrule(lr){6-7}
\cmidrule(lr){9-10}
&
\textbf{Dev} &
\textbf{Test} &
\textbf{zh} &
\textbf{en} &
\textbf{zh} &
\textbf{en} &
\textbf{test} &
\textbf{net} &
\textbf{meeting} &
\textbf{en} \\
\midrule
\rowcolor{headerblue}
\multicolumn{11}{l}{\textit{Closed-source models}} \\
Gemini-3-Flash &
1.7|3.56 &
1.81|4.91 &
13.58 &
8.49 &
7.52 &
4.01 &
2.66 &
14.38 &
17.62 &
7.74 \\
Doubao-LLM ASR &
2.95|4.06 &
2.92|5.32 &
\textbf{4.60} &
7.12 &
2.92 &
7.22 &
0.98 &
\textbf{4.46} &
\textbf{4.90} &
7.14 \\
GPT-4o-trans. &
1.52|3.29 &
1.75|4.23 &
12.61 &
7.22 &
2.62 &
\textbf{2.71} &
3.52 &
15.71 &
31.40 &
7.02 \\
\rowcolor{headerblue}
\multicolumn{11}{l}{\textit{Open-source models}} \\
Canary-1B-v2 &
2.07|4.03 &
2.20|3.58 &
- &
8.91 &
- &
4.48 &
- &
- &
- &
6.20 \\
Parakeet-TDT-0.6B-v3 &
1.91|3.54 &
1.93|3.60 &
- &
8.54 &
- &
4.88 &
- &
- &
- &
6.11 \\
Voxtral-Mini-3B-2507 &
1.89|3.88 &
1.89|4.08 &
- &
10.15 &
- &
3.84 &
- &
- &
- &
7.08 \\
Step-Audio-2-mini &
\textbf{1.21}|\textbf{2.50} &
1.37|2.75 &
\underline{4.77} &
\textbf{7.04} &
\textbf{2.48} &
3.93 &
0.81 &
5.56 &
\underline{5.46} &
7.43 \\
Kimi-Audio-7B &
1.38|2.56 &
\textbf{1.34}|\textbf{2.55} &
6.74 &
8.35 &
5.88 &
8.07 &
\textbf{0.76} &
6.41 &
6.25 &
8.15 \\
Whisper Large-v3 &
1.74|3.68 &
1.78|3.53 &
15.33 &
16.18 &
7.70 &
4.10 &
5.89 &
12.02 &
17.79 &
9.00 \\
Qwen2.5-Omni-7B &
2.05|4.19 &
2.37|4.21 &
5.01 &
8.56 &
4.64 &
4.01 &
1.15 &
6.16 &
9.64 &
\textbf{6.02} \\
Qwen3-ASR-1.7B &
1.62|3.07 &
1.62|3.40 &
7.42 &
7.57 &
3.93 &
\underline{3.19} &
1.52 &
4.99 &
5.80 &
6.25 \\
\rowcolor{headerblue}
\multicolumn{11}{l}{\textit{Our model}} \\
Ours &
1.62|3.21 &
1.78|3.57 &
5.8 &
8.15 &
5.43 &
3.76 &
1.49 &
5.19 &
6.17 &
7.44 \\
Ours w/ router &
1.64|3.07 &
1.63|3.37 &
7.37 &
7.57 &
3.86 &
3.17 &
1.53 &
\underline{4.95} &
5.89 &
6.26 \\
\bottomrule
\end{tabular}

\vspace{-5mm}
\end{table*}

\vspace*{-6mm}
\section{\vspace{-2mm}Experiments}
\vspace{-1mm}
\subsection{Experimental setup}
\vspace{-2mm}

\paragraph{Datasets and Evaluation.}
We initialize from \textbf{Qwen3-ASR-1.7B}~\citep{qwen3-asr} and train on \textbf{ \textsc{Voices-in-the-wild-2M} } for both SFT and RL 
stages. We evaluate along three axes. \textit{(i) Standard ASR}: 
LibriSpeech~\citep{librispeech}, CommonVoice22~\citep{commonvoice}, FLEURS~\citep{fleurs}, 
AISHELL-1~\citep{aishell}, WenetSpeech~\citep{wenetspeech}, and VoxPopuli~\citep{voxpopuli}, 
reported with and without our dynamic routing LoRA to verify that 
robustness adaptation does not regress clean-speech performance. 
\textit{(ii) Adverse-condition ASR}: CHiME-4~\citep{CHIME4}, 
VOiCES~\citep{VOICES}, and NOIZEUS~\citep{Noizeus}, covering noise, reverberation, 
far-field, and signal degradation. \textit{(iii) Compound conditions}: 
our \textbf{Voices-in-the-Wild-Bench}, targeting realistic multi-factor 
acoustic environments.

\vspace{-2mm}

\paragraph{Baselines.}
We compare against 12 representative systems spanning conventional ASR, 
large audio-language models, and omni-modal foundation models: 
Whisper-Large-v3~\citep{whisper}, Canary-1B-v2~\citep{canary&parakeet}, 
Parakeet-TDT-0.6B-v3~\citep{canary&parakeet}, 
Qwen2.5-Omni-7B~\citep{qwen2.5omni}, Step-Audio-2-mini~\citep{stepaudio2}, Voxtral-Mini-3B~\citep{voxtral}, Kimi-Audio-7B~\citep{Kimi-Audio}, Gemini-3-Flash~\citep{}, Seed-ASR~\citep{seedasr}, 
GPT-4o~\citep{gpt-4o}, and Step-Audio-2~\citep{stepaudio2}.

\paragraph{Implementation Details.}
A2S-SFT uses learning rates of $1{\times}10^{-3}$ for the audio encoder 
and adapter, $2{\times}10^{-5}$ for the LLM, and $2{\times}10^{-6}$ for 
the joint stage. RL runs for 6{,}000 steps with learning rate 
$1{\times}10^{-6}$ and $K{=}16$ rollouts per input, optimized under the 
combined reward  $0.4\,R_\text{rule} + 0.6\,R_\text{dynamic}$. 

\subsection{Main results}
The main results demonstrate \textbf{3} key findings, verifying that \textsc{Mega-ASR} achieves strong robustness from clean speech to highly compositional real-world acoustic environments. \textbf{[Enh.1] Competitive general ASR with adaptive routing (Table~\ref{tab:main_standard}).} \textsc{Mega-ASR} remains highly competitive on clean and multilingual benchmarks against Qwen3-ASR, Seed-ASR, and Kimi-Audio. With routing, it improves LibriSpeech WER from 1.78/3.57 to 1.63/3.37, achieves 3.86/3.17 on Fleurs zh/en, and shows consistent gains on WenetSpeech-meeting and VoxPopuli. \textbf{[Enh.2] State-of-the-art robustness under acoustic perturbations (Table~\ref{tab:main_noise} Figure~\ref{fig:radar_comparison}).} \textsc{Mega-ASR} achieves the best overall robustness on CHiME-4, VOiCES, and NOIZEUS with an average WER of 6.70, outperforming Qwen3-ASR (7.93), Whisper-Large-v3 (10.72), and Qwen2.5-Omni (15.14). Under extreme NOIZEUS 0dB conditions, it further reduces WER to 19.80 versus 23.97 for Qwen3-ASR and 55.78 for Gemini-3-Flash, a relative reduction of 17.4\% over the strongest baseline and 64.5\% over Gemini-3-Flash. \textbf{[Enh.3] Superior robustness in compositional real-world environments (Table~\ref{tab:world_breakdown}).} On Voices-in-the-Wild-Bench, \textsc{Mega-ASR} consistently achieves the strongest performance across mixed degradations, far-field speech, and recording artifacts. Under mixed degradations, it achieves 2.73/4.57 WER, substantially outperforming Whisper-Large-v3 (8.91/14.79) and Gemini-3-Flash (7.99/9.62), corresponding to a 65.8\%/69.1\% relative reduction over Whisper-Large-v3 and 65.8\% over Gemini-3-Flash.

\begin{table*}[t]
\centering
\caption{Breakdown results on \textsc{Voices-in-the-Wild-Bench} by acoustic scenario.}
\vspace{-1mm}
\label{tab:world_breakdown}
{\footnotesize
\setlength{\tabcolsep}{2pt}
\begin{tabular}{lcccccccccccccccc}
\toprule
\multirow{2}{*}{\textbf{Model}}
& \multicolumn{2}{c}{\textbf{Noise}}
& \multicolumn{2}{c}{\textbf{Far.}}
& \multicolumn{2}{c}{\textbf{Obst.}}
& \multicolumn{2}{c}{\textbf{Echo.}}
& \multicolumn{2}{c}{\textbf{Record.}}
& \multicolumn{2}{c}{\textbf{Elc.Dis.}}
& \multicolumn{2}{c}{\textbf{Trans.Drop.}}
& \multicolumn{2}{c}{\textbf{Mixed}} \\
\cmidrule(lr){2-3}
\cmidrule(lr){4-5}
\cmidrule(lr){6-7}
\cmidrule(lr){8-9}
\cmidrule(lr){10-11}
\cmidrule(lr){12-13}
\cmidrule(lr){14-15}
\cmidrule(lr){16-17}
& \textbf{Real.}
& \textbf{Sim.}
& \textbf{Real.}
& \textbf{Sim.}
& \textbf{Real.}
& \textbf{Sim.}
& \textbf{Real.}
& \textbf{Sim.}
& \textbf{Real.}
& \textbf{Sim.}
& \textbf{Real.}
& \textbf{Sim.}
& \textbf{Real.}
& \textbf{Sim.}
& \textbf{Real.}
& \textbf{Sim.} \\
\midrule
\rowcolor{headerblue}
\multicolumn{17}{l}{\textit{Closed-source models}} \\
Gemini3-Flash
& 7.63 & 10.61 & 5.14 & 1.90 & 3.73 & 2.65 & 8.75 & 14.86 & 8.38 & 19.85 & 3.15 & 7.56 & 5.47 & 7.65 & 7.99 & 9.62 \\
Seed-ASR
& 8.21 & 8.11 & 3.06 & 3.19 & 3.10 & 2.76 & 16.55 & 18.21 & 18.48 & 23.33 & 3.89 & 5.71 & 7.97 & 7.46 & 6.88 & 9.29 \\
GPT-4o-trans.
& 13.19 & 45.78 & \textbf{1.87} & 2.39 & \textbf{1.57} & 2.77 & 15.62 & 28.76 & 13.37 & 22.60 & 3.70 & 8.43 & 8.76 & 7.71 & 5.62 & 11.00 \\
\rowcolor{headerblue}
\multicolumn{17}{l}{\textit{Open-source models}} \\
Whisper-L-v3
& 16.57 & 18.19 & 3.38 & 6.85 & 3.06 & 6.01 & 25.34 & 39.87 & 18.33 & 31.81 & 3.74 & 8.77 & 7.04 & 8.05 & 8.91 & 14.79 \\
Qwen2.5-Omni
& 11.92 & 17.88 & 2.35 & 2.44 & 2.40 & 2.08 & 20.01 & 32.64 & 13.71 & 30.09 & 2.46 & 5.96 & 6.34 & 5.88 & 6.40 & 10.29 \\
Kimi-Audio
& 35.10 & 14.59 & 2.71 & 1.92 & 2.49 & 1.64 & 24.00 & 26.58 & 8.73 & 18.09 & 1.83 & \textbf{2.78} & 4.54 & 6.33 & 4.44 & 6.19 \\
Qwen3-ASR
& 7.51 & 9.52 & \underline{2.23} & \textbf{1.54} & 1.73 & \underline{1.27} & 10.40 & 14.61 & 9.57 & 19.42 & \textbf{1.54} & 3.41 & 4.16 & 4.19 & 3.30 & 5.39 \\
\rowcolor{headerblue}
\multicolumn{17}{l}{\textit{Our model}} \\
Ours
& \underline{6.33} & \underline{8.26} & 2.35 & \underline{1.61} & \underline{1.62} & \textbf{1.23} & \textbf{8.62} & \underline{12.59} & \underline{7.65} & \underline{14.21} & 1.71 & 3.72 & \textbf{2.59} & \textbf{2.62} & \underline{2.73} & \underline{4.57} \\
Ours w/ router
& \textbf{6.12} & \textbf{8.09} & 2.33 & 1.69 & 1.80 & 1.41 & \underline{8.66} & \textbf{12.22} & \textbf{6.91} & \textbf{13.23} & \underline{1.60} & \underline{3.35} & \underline{2.72} & \underline{2.88} & \textbf{2.63} & \textbf{4.53} \\
\bottomrule
\end{tabular}
}
\end{table*}

\vspace{-2mm}
\subsection{Analysis}
\vspace{-2mm}

\begin{figure*}[!h]
\begin{minipage}[c]{0.46\textwidth}
Through ablation studies, we derive five key observations (\textbf{[Obs.1]}--\textbf{[Obs.5]}) spanning semantic-level gains, training recipe, reward design, and hyperparameter sensitivity. We elaborate each below, with the corresponding evidence drawn from Tables~\ref{tab:ablation}--\ref{tab:hp-tau} \paragraph{\textbf{[Obs.1] Mega-ASR's gains generalize beyond WER to semantic-level metrics.}}
Table~\ref{tab:judge} shows consistent semantic-level improvements over Qwen3-ASR, with missed-content dropping from $14.2$ to $5.9$. This validates that \textsc{Mega-ASR} delivers semantic- and holistic-level gains, exemplified by reduced hallucination and dropped utterances, beyond merely lowering WER.
\end{minipage}%
\hfill
\begin{minipage}[c]{0.52\textwidth}
\centering
\captionof{table}{A2S-SFT and DG-WGPO ablation. WER (\%, $\downarrow$) on Voices/Noizeus mid+high.}
\label{tab:ablation}
\small
\begin{tabular}{lcc}
\toprule
\textbf{Variant} & \textbf{Voices} & \textbf{Noizeus} \\
\midrule
Qwen3-ASR (baseline)            & 8.94 & 9.45 \\
\midrule
+ SFT w/o A2S     & 8.31 & 8.79 \\
Mega-ASR-Base    &  7.59 & 8.12 \\
\midrule
+ vanilla GRPO ($R_\text{wer}$ only) & 7.73 & 8.11 \\
+ vanilla DAPO ($R_\text{wer}$ only) & 7.62 & 7.98 \\
+ DG-WGPO w/o $R_\text{rep}$    & 7.46 & 7.73 \\
+ DG-WGPO w/o $R_\text{fine}$   & 7.45 & 7.71 \\
+ DG-WGPO w/o $R_\text{struc}$  & 7.54 & 7.85 \\
+ DG-WGPO w/o gated fusion      & 7.41 & 7.68 \\
\midrule
\textbf{Mega-ASR (full)}        & \textbf{7.35} & \textbf{7.64} \\
\bottomrule
\end{tabular}
\end{minipage}
\end{figure*}

\vspace{-2mm}
\begin{table}[t]
\centering
\small
\setlength{\tabcolsep}{3.8pt}
\renewcommand{\arraystretch}{0.92}
\begin{minipage}[t]{0.47\linewidth}
\centering
\captionof{table}{Reward design. WER (\%, $\downarrow$) on three test sets and average training time per step (Avg.\ T., relative).}
\label{tab:rm}
\begin{tabular}{lcccc}
\toprule
\textbf{Reward} & \textbf{Voices} & \textbf{Noizeus} & \textbf{Voi-R.} & \textbf{Avg.T.} \\
\midrule
LLM-judge           & 7.51 & 7.71 & 9.27 &  62.23\\
\textbf{Rule-based} & \textbf{7.53} & \textbf{7.64} & \textbf{9.38} & \textbf{19.57} \\
\bottomrule
\end{tabular}
\vspace{1mm}
\captionof{table}{LLM-as-judge evaluation. Avg over Voices and Noizeus.}
\label{tab:judge}
\vspace{-2mm}
\begin{tabular}{lcccc}
\toprule
\textbf{Model} & \textbf{Hall.} & \textbf{Miss} & \textbf{Sem.} & \textbf{KeyE.} \\
\midrule
Qwen3-ASR        & 18.7 & 14.2 & 71.3 & 22.5 \\
Mega-ASR-Base    & 15.4 & 11.6 & 79.8 & 20.1 \\
\textbf{Mega-ASR}& \textbf{11.8} & \textbf{5.9} & \textbf{86.4} & \textbf{19.5} \\
\bottomrule
\end{tabular}
\end{minipage}
\hfill
\begin{minipage}[t]{0.5\linewidth}
\centering
\setlength{\tabcolsep}{4.5pt}
\renewcommand{\arraystretch}{1.0}
\small
\vspace{0.5mm}
\captionof{table}{Sensitivity to reward weights $(\alpha_{\text{dyn}},\alpha_s)$. WER (\%, $\downarrow$) is reported on four held-out subsets grouped by degradation type (V.N.R.: Voices-Noise-Real; V.F.R.: Voices-Far-Real).}
\vspace{2mm}
\label{tab:hp-alpha}
\begin{tabular}{lcccc}
\toprule
\multirow{2}{*}{\textbf{Settings}}  & \multicolumn{2}{c}{\textbf{Noise}} & \multicolumn{2}{c}{\textbf{Far}} \\
\cmidrule(lr){2-3} \cmidrule(lr){4-5}
 & Nz & V.N.R. & V.F. & V.F.R. \\
\midrule
$\alpha_{\text{dyn}}{=}0.4,\ \alpha_s{=}0.4$ & 7.7 & 7.6 & 7.8 & 9.5 \\
$\alpha_{\text{dyn}}{=}0.4,\ \alpha_s{=}0.6$ & 7.8 & 7.6 & 7.9 & 9.4 \\
$\alpha_{\text{dyn}}{=}0.6,\ \alpha_s{=}0.2$ & 7.8 & 7.5 & 7.6 & 9.3 \\
$\alpha_{\text{dyn}}{=}0.6,\ \alpha_s{=}0.6$ & \textbf{7.5} & 7.5 & 7.4 & 9.3 \\
$\alpha_{\text{dyn}}{=}0.8,\ \alpha_s{=}0.4$ & 8.1 & 9.1 & 8.0 & 9.9 \\
$\boldsymbol{\alpha_{\text{dyn}}{=}0.6,\ \alpha_s{=}0.4}$ & 7.6 & \textbf{7.4} & \textbf{7.4} & \textbf{9.2} \\
\bottomrule
\end{tabular}
\end{minipage}
\vspace{-3mm}
\end{table}

\vspace{-2mm}

\paragraph{\textbf{[Obs.2] Ablation of A2S-SFT and DG-WGPO components.}}
We ablate each stage of A2S-SFT and each component of DG-WGPO on Voices/Noizeus in Table~\ref{tab:ablation}. Removing the first two progressive stages (SFT w/o A2S) reaches $8.31/8.79$ WER, still $0.72/0.67$ behind Mega-ASR-Base, confirming the value of staged acoustic-to-semantic adaptation. On top of Mega-ASR-Base, vanilla DAPO with $R_{\text{wer}}$ alone outperforms vanilla GRPO by $0.11/0.13$ WER, motivating our choice of DAPO as the RL backbone. Among the DG-WGPO components, removing $R_{\text{struc}}$ causes the largest degradation ($7.54/7.85$), indicating that sentence-level reconstruction is critical on mid- and high-WER samples; removing $R_{\text{rep}}$, $R_{\text{fine}}$, or gated fusion each yields a smaller but consistent drop. The full \textsc{Mega-ASR} reaches $7.35/7.64$, a $1.59/1.81$ reduction over Qwen3-ASR.

\vspace{-2mm}
\paragraph{\textbf{[Obs.3] Rule-based reward matches LLM-judge at $3.2\times$ lower time-cost.}}
We replace $R_\text{dynamic}$ with a Gemini-2.5-flash-lite scalar score and compare it against our rule-based design (Table~\ref{tab:rm}). The two variants achieve comparable WER across all three test sets, with differences within roughly 0.1 on Voices and Noizeus and 0.11 on Voi-R., suggesting that the rule-based reward already captures the supervision signals an LLM judge would provide. The LLM-judge variant, however, takes 62.23s per training step compared to 19.57s for the rule-based reward, a $3.2\times$ slowdown that scales unfavorably with longer training. Given the negligible accuracy difference and the substantial computational overhead, we adopt the rule-based design as the default.

\vspace{-2mm}
\paragraph{\textbf{[Obs.4] Ablation on hyperparameters.}}
We perturb $\alpha_{\text{dyn}}$ and $\alpha_s$ around the default $(0.6, 0.4)$ in Table~\ref{tab:hp-alpha}. Pushing $\alpha_{\text{dyn}}$ to 0.8 causes the sharpest degradation, with V.N.R.\ rising from 7.4 to 9.1 and Nz from 7.6 to 8.1, indicating that an over-weighted gating term suppresses the dominant WER-driven signal $R_{\text{wer}}$ and harms recognition. Lowering $\alpha_{\text{dyn}}$ to 0.4 instead hurts the far-field subsets, where V.F.\ rises by 0.4 and V.F.R.\ by 0.3, while varying $\alpha_s$ in $\{0.2, 0.6\}$ produces only minor fluctuations across all four subsets. These observations suggest that $\alpha_{\text{dyn}}$ governs a more sensitive trade-off than $\alpha_s$, and we therefore adopt $(\alpha_{\text{dyn}}, \alpha_s){=}(0.6, 0.4)$, which achieves the best or near-best WER on every subset.

\vspace{-2mm}
\begin{minipage}[c]{0.55\textwidth}
We further sweep the gating threshold $\tau$ from 0.2 to 0.5 (Table~\ref{tab:hp-tau}). The trend mirrors our earlier observation: $\tau{=}0.3$ gives the most balanced result, $\tau{=}0.2$ and $\tau{=}0.4$ have only marginal effect, while $\tau{=}0.5$ leads to a clear degradation, consistent with the over-restrictive gating effect seen at high $\alpha_{\text{dyn}}$.
\end{minipage}%
\hfill
\begin{minipage}[c]{0.42\textwidth}
\centering
\vspace{-2mm}
\captionof{table}{Sensitivity to gating threshold $\tau$. WER (\%, $\downarrow$) on Noizeus.}
\label{tab:hp-tau}
\small
\setlength{\tabcolsep}{6pt}
\begin{tabular}{ccccc}
\toprule
$\tau$ & 0.2 & \textbf{0.3} & 0.4 & 0.5 \\
\midrule
Noizeus & 7.68 & \textbf{7.64} & 7.66 & 7.70 \\
\bottomrule
\end{tabular}
\end{minipage}

\vspace{-2mm}
\section{Case study}
\begin{figure}[!t]
    \centering
    \includegraphics[width=1.01\linewidth]{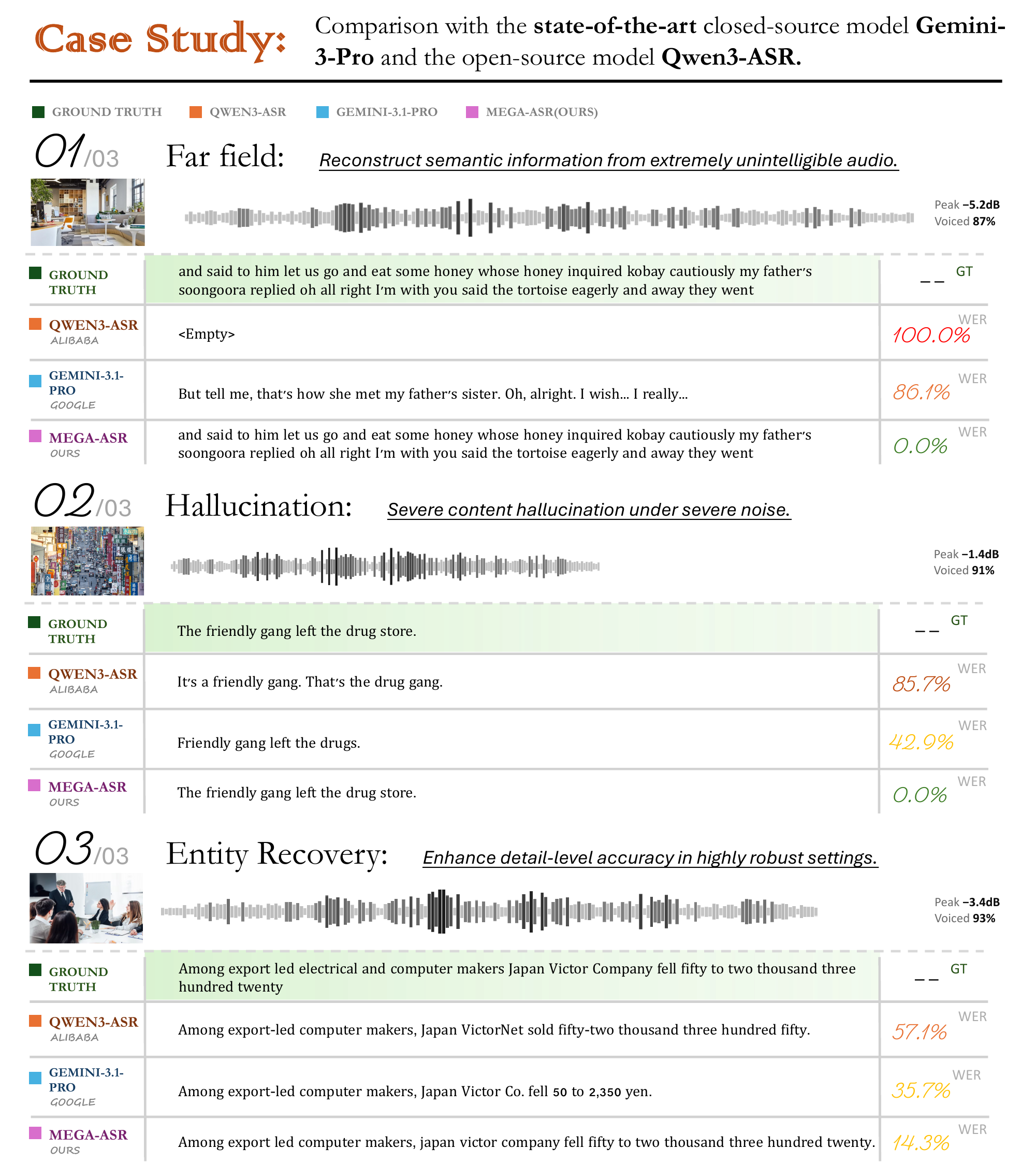}
    \caption{Case study against SOTA models \textsc{Gemini-3-Pro} and \textsc{Qwen3-ASR} on semantic reconstruction under strong environmental robustness, hallucination, and fine-grained detail recovery. \textsc{Mega-ASR} faithfully aligns with the reference transcript (WER $\mathbf{0.0\%}$ on far-field), while competing SOTA systems either return empty outputs or fabricate fluent but incorrect content.}
    \label{fig:study}
\end{figure}

\vspace{-2mm}
Figure~\ref{fig:study} presents a comparative case study where the state-of-the-art closed-source model \textsc{Gemini-3-Pro}, the open-source model \textsc{Qwen3-ASR}, and our proposed \textsc{Mega-ASR} transcribe the same challenging audio across three scenarios: far-field reconstruction, content hallucination, and entity recovery. In the far-field case (Peak \textbf{-5.2 dB}), \textsc{Qwen3-ASR} offers only a superficial response, returning an empty transcription with a WER of \textbf{100.0\%}. \textsc{Gemini-3-Pro} goes beyond this and produces a fluent hypothesis, yet fabricates content unrelated to the source (WER \textbf{86.1\%}). In contrast, \textsc{Mega-ASR} precisely recovers the reference transcript (WER \textbf{0.0\%}), a pattern that persists under severe noise and in entity-dense utterances. This highlights the intrinsic difficulty of robust speech recognition: errors are often subtle, originate from degraded signals or rare entities, and remain hidden behind outputs that appear fluent at the surface level.

\vspace{-2mm}
\section{Conclusion}
\vspace{-2mm}
We presented \textbf{MEGA-ASR}, a unified ASR-in-the-wild framework designed to overcome the acoustic robustness bottleneck of current ASR and large audio-language models under severe, compositional distortions. Central to MEGA-ASR is \textbf{VOICES-IN-THE-WILD-2M}, a large-scale dataset covering \textbf{7} classic acoustic phenomena and \textbf{54} physically plausible compound scenarios, together with \textbf{Acoustic-to-Semantic Progressive Supervised Fine-Tuning} and \textbf{Dual-Granularity WER-Gated Policy Optimization} for robust perceptual recovery and semantic reconstruction. Extensive experiments show that MEGA-ASR achieves significant improvements over prior state-of-the-art systems, especially under challenging real-world acoustic conditions where relative WER reductions can exceed \(30\%\). Our results highlight the importance of modeling compound acoustic environments at scale and establish MEGA-ASR as a scalable paradigm for robust ASR in-the-wild.

\bibliographystyle{plainnat}   
\bibliography{main}

@article{qwen3-asr,
  title={Qwen3-ASR Technical Report},
  author={Shi, Xian and Wang, Xiong and Guo, Zhifang and Wang, Yongqi and Zhang, Pei and Zhang, Xinyu and Guo, Zishan and Hao, Hongkun and Xi, Yu and Yang, Baosong and others},
  journal={arXiv preprint arXiv:2601.21337},
  year={2026}
}

@article{fireredasr2s,
  title={FireRedASR2S: A State-of-the-Art Industrial-Grade All-in-One Automatic Speech Recognition System},
  author={Xu, Kaituo and Jia, Yan and Huang, Kai and Chen, Junjie and Li, Wenpeng and Liu, Kun and Xie, Feng-Long and Tang, Xu and Hu, Yao},
  journal={arXiv preprint arXiv:2603.10420},
  year={2026}
}

@article{mini-omni-reasoner,
  title={Mini-omni-reasoner: Token-level thinking-in-speaking in large speech models},
  author={Xie, Zhifei and Ma, Ziyang and Liu, Zihang and Pang, Kaiyu and Li, Hongyu and Zhang, Jialin and Liao, Yue and Ye, Deheng and Miao, Chunyan and Yan, Shuicheng},
  journal={arXiv preprint arXiv:2508.15827},
  year={2025}
}

@article{qwen3-omni,
  title={Qwen3-omni technical report},
  author={Xu, Jin and Guo, Zhifang and Hu, Hangrui and Chu, Yunfei and Wang, Xiong and He, Jinzheng and Wang, Yuxuan and Shi, Xian and He, Ting and Zhu, Xinfa and others},
  journal={arXiv preprint arXiv:2509.17765},
  year={2025}
}

@article{Kimi-Audio,
  title={Kimi-audio technical report},
  author={Ding, Ding and Ju, Zeqian and Leng, Yichong and Liu, Songxiang and Liu, Tong and Shang, Zeyu and Shen, Kai and Song, Wei and Tan, Xu and Tang, Heyi and others},
  journal={arXiv preprint arXiv:2504.18425},
  year={2025}
}

@article{Noizeus,
  title={Subjective comparison and evaluation of speech enhancement algorithms},
  author={Hu, Yi and Loizou, Philipos C},
  journal={Speech communication},
  volume={49},
  number={7-8},
  pages={588--601},
  year={2007},
  publisher={Elsevier}
}

@inproceedings{ted-lium,
  title={TED-LIUM: an automatic speech recognition dedicated corpus.},
  author={Rousseau, Anthony and Del{\'e}glise, Paul and Esteve, Yannick},
  booktitle={LREC},
  pages={125--129},
  year={2012}
}

@article{VOICES,
  title={Voices obscured in complex environmental settings (voices) corpus},
  author={Richey, Colleen and Barrios, Maria A and Armstrong, Zeb and Bartels, Chris and Franco, Horacio and Graciarena, Martin and Lawson, Aaron and Nandwana, Mahesh Kumar and Stauffer, Allen and van Hout, Julien and others},
  journal={arXiv preprint arXiv:1804.05053},
  year={2018}
}

@article{BERSt,
  title={BERSting at the screams: a benchmark for distanced, emotional and shouted speech recognition},
  author={Tutt{\"o}s{\'\i}, Paige and Dhillon, Mantaj and Sang, Luna and Eastwood, Shane and Bhatia, Poorvi and Dinh, Quang Minh and Kapoor, Avni and Jin, Yewon and Lim, Angelica},
  journal={Computer Speech \& Language},
  volume={95},
  pages={101815},
  year={2026},
  publisher={Elsevier}
}

@article{DAPS,
  title={Can we automatically transform speech recorded on common consumer devices in real-world environments into professional production quality speech?—a dataset, insights, and challenges},
  author={Mysore, Gautham J},
  journal={IEEE Signal Processing Letters},
  volume={22},
  number={8},
  pages={1006--1010},
  year={2014},
  publisher={IEEE}
}

@inproceedings{librispeech,
  title={Librispeech: an asr corpus based on public domain audio books},
  author={Panayotov, Vassil and Chen, Guoguo and Povey, Daniel and Khudanpur, Sanjeev},
  booktitle={2015 IEEE international conference on acoustics, speech and signal processing (ICASSP)},
  pages={5206--5210},
  year={2015},
  organization={IEEE}
}

@inproceedings{commonvoice,
  title={Common voice: A massively-multilingual speech corpus},
  author={Ardila, Rosana and Branson, Megan and Davis, Kelly and Kohler, Michael and Meyer, Josh and Henretty, Michael and Morais, Reuben and Saunders, Lindsay and Tyers, Francis and Weber, Gregor},
  booktitle={Proceedings of the twelfth language resources and evaluation conference},
  pages={4218--4222},
  year={2020}
}

@inproceedings{wenetspeech,
  title={Wenetspeech: A 10000+ hours multi-domain mandarin corpus for speech recognition},
  author={Zhang, Binbin and Lv, Hang and Guo, Pengcheng and Shao, Qijie and Yang, Chao and Xie, Lei and Xu, Xin and Bu, Hui and Chen, Xiaoyu and Zeng, Chenchen and others},
  booktitle={ICASSP 2022-2022 IEEE International Conference on Acoustics, Speech and Signal Processing (ICASSP)},
  pages={6182--6186},
  year={2022},
  organization={IEEE}
}

@inproceedings{aishell,
  title={Aishell-1: An open-source mandarin speech corpus and a speech recognition baseline},
  author={Bu, Hui and Du, Jiayu and Na, Xingyu and Wu, Bengu and Zheng, Hao},
  booktitle={2017 20th conference of the oriental chapter of the international coordinating committee on speech databases and speech I/O systems and assessment (O-COCOSDA)},
  pages={1--5},
  year={2017},
  organization={IEEE}
}

@article{MUSAN,
  title={Musan: A music, speech, and noise corpus},
  author={Snyder, David and Chen, Guoguo and Povey, Daniel},
  journal={arXiv preprint arXiv:1510.08484},
  year={2015}
}

@article{DNS,
  title={The interspeech 2020 deep noise suppression challenge: Datasets, subjective testing framework, and challenge results},
  author={Reddy, Chandan KA and Gopal, Vishak and Cutler, Ross and Beyrami, Ebrahim and Cheng, Roger and Dubey, Harishchandra and Matusevych, Sergiy and Aichner, Robert and Aazami, Ashkan and Braun, Sebastian and others},
  journal={arXiv preprint arXiv:2005.13981},
  year={2020}
}

@inproceedings{esc-50,
  title={ESC: Dataset for environmental sound classification},
  author={Piczak, Karol J},
  booktitle={Proceedings of the 23rd ACM international conference on Multimedia},
  pages={1015--1018},
  year={2015}
}

@inproceedings{URBANSOUND,
  title={A dataset and taxonomy for urban sound research},
  author={Salamon, Justin and Jacoby, Christopher and Bello, Juan Pablo},
  booktitle={Proceedings of the 22nd ACM international conference on Multimedia},
  pages={1041--1044},
  year={2014}
}

@inproceedings{fleurs,
  title={Fleurs: Few-shot learning evaluation of universal representations of speech},
  author={Conneau, Alexis and Ma, Min and Khanuja, Simran and Zhang, Yu and Axelrod, Vera and Dalmia, Siddharth and Riesa, Jason and Rivera, Clara and Bapna, Ankur},
  booktitle={2022 IEEE Spoken Language Technology Workshop (SLT)},
  pages={798--805},
  year={2023},
  organization={IEEE}
}

@article{voxpopuli,
  title={Crowdspeech and voxdiy: Benchmark datasets for crowdsourced audio transcription},
  author={Pavlichenko, Nikita and Stelmakh, Ivan and Ustalov, Dmitry},
  journal={arXiv preprint arXiv:2107.01091},
  year={2021}
}

@article{CHIME4,
  title={The 4th CHiME speech separation and recognition challenge},
  author={Watanabe, E Vincent S and Mandel, Michael and Barker, Jon},
  year={2016}
}

@article{paraformer,
  title={Paraformer: Fast and accurate parallel transformer for non-autoregressive end-to-end speech recognition},
  author={Gao, Zhifu and Zhang, Shiliang and McLoughlin, Ian and Yan, Zhijie},
  journal={arXiv preprint arXiv:2206.08317},
  year={2022}
}

@inproceedings{reasoningforasr,
  title={Error correction for speech recognition systems using large language model reasoning capabilities},
  author={Lina, Sun and Aksyonov, Konstantin A},
  booktitle={2024 IEEE 25th International Conference of Young Professionals in Electron Devices and Materials (EDM)},
  pages={2300--2303},
  year={2024},
  organization={IEEE}
}

@inproceedings{ko2015audio,
  title={Audio augmentation for speech recognition.},
  author={Ko, Tom and Peddinti, Vijayaditya and Povey, Daniel and Khudanpur, Sanjeev},
  booktitle={Interspeech},
  volume={2015},
  pages={3586},
  year={2015}
}

@inproceedings{ko2017audio,
  title={A study on data augmentation of reverberant speech for robust speech recognition},
  author={Ko, Tom and Peddinti, Vijayaditya and Povey, Daniel and Seltzer, Michael L and Khudanpur, Sanjeev},
  booktitle={2017 IEEE international conference on acoustics, speech and signal processing (ICASSP)},
  pages={5220--5224},
  year={2017},
  organization={IEEE}
}

@article{parada2022pmctpatchedmulticonditiontraining,
  title={pMCT: Patched multi-condition training for robust speech recognition},
  author={Parada, Pablo Peso and Dobrowolska, Agnieszka and Saravanan, Karthikeyan and Ozay, Mete},
  journal={arXiv preprint arXiv:2207.04949},
  year={2022}
}

@inproceedings{yan2025improving,
  title={Improving multilingual ASR in the wild using simple N-best re-ranking},
  author={Yan, Brian and Pratap, Vineel and Watanabe, Shinji and Auli, Michael},
  booktitle={ICASSP 2025-2025 IEEE International Conference on Acoustics, Speech and Signal Processing (ICASSP)},
  pages={1--5},
  year={2025},
  organization={IEEE}
}

@inproceedings{asrinthewild,
  title={Deep Learning-Based Telephony Speech Recognition in the Wild.},
  author={Han, Kyu Jeong and Hahm, Seongjun and Kim, Byung-Hak and Kim, Jungsuk and Lane, Ian R},
  booktitle={Interspeech},
  pages={1323--1327},
  year={2017}
}

@article{mini-omni2,
  title={Mini-omni2: Towards open-source gpt-4o with vision, speech and duplex capabilities},
  author={Xie, Zhifei and Wu, Changqiao},
  journal={arXiv preprint arXiv:2410.11190},
  year={2024}
}

@article{dapo,
  title={Dapo: An open-source llm reinforcement learning system at scale},
  author={Yu, Qiying and Zhang, Zheng and Zhu, Ruofei and Yuan, Yufeng and Zuo, Xiaochen and Yue, Yu and Dai, Weinan and Fan, Tiantian and Liu, Gaohong and Liu, Lingjun and others},
  journal={arXiv preprint arXiv:2503.14476},
  year={2025}
}

@inproceedings{whisper,
  title={Robust speech recognition via large-scale weak supervision},
  author={Radford, Alec and Kim, Jong Wook and Xu, Tao and Brockman, Greg and McLeavey, Christine and Sutskever, Ilya},
  booktitle={International conference on machine learning},
  pages={28492--28518},
  year={2023},
  organization={PMLR}
}

@article{funasr,
  title={Funasr: A fundamental end-to-end speech recognition toolkit},
  author={Gao, Zhifu and Li, Zerui and Wang, Jiaming and Luo, Haoneng and Shi, Xian and Chen, Mengzhe and Li, Yabin and Zuo, Lingyun and Du, Zhihao and Xiao, Zhangyu and others},
  journal={arXiv preprint arXiv:2305.11013},
  year={2023}
}

@misc{qwen2.5omni,
      title={Qwen2.5-Omni Technical Report}, 
      author={Jin Xu and Zhifang Guo and Jinzheng He and Hangrui Hu and Ting He and Shuai Bai and Keqin Chen and Jialin Wang and Yang Fan and Kai Dang and Bin Zhang and Xiong Wang and Yunfei Chu and Junyang Lin},
      year={2025},
      eprint={2503.20215},
      archivePrefix={arXiv},
      primaryClass={cs.CL},
      url={https://arxiv.org/abs/2503.20215}, 
}

@article{stepaudio2,
  title={Step-audio 2 technical report},
  author={Wu, Boyong and Yan, Chao and Hu, Chen and Yi, Cheng and Feng, Chengli and Tian, Fei and Shen, Feiyu and Yu, Gang and Zhang, Haoyang and Li, Jingbei and others},
  journal={arXiv preprint arXiv:2507.16632},
  year={2025}
}

@article{canary&parakeet,
  title={Canary-1b-v2 \& parakeet-tdt-0.6 b-v3: Efficient and high-performance models for multilingual asr and ast},
  author={Sekoyan, Monica and Koluguri, Nithin Rao and Tadevosyan, Nune and Zelasko, Piotr and Bartley, Travis and Karpov, Nikolay and Balam, Jagadeesh and Ginsburg, Boris},
  journal={arXiv preprint arXiv:2509.14128},
  year={2025}
}

@article{voxtral,
  title={Voxtral},
  author={Liu, Alexander H and Ehrenberg, Andy and Lo, Andy and Denoix, Cl{\'e}ment and Barreau, Corentin and Lample, Guillaume and Delignon, Jean-Malo and Chandu, Khyathi Raghavi and von Platen, Patrick and Muddireddy, Pavankumar Reddy and others},
  journal={arXiv preprint arXiv:2507.13264},
  year={2025}
}

@article{seedasr,
  title={Seed-asr: Understanding diverse speech and contexts with llm-based speech recognition},
  author={Bai, Ye and Chen, Jingping and Chen, Jitong and Chen, Wei and Chen, Zhuo and Ding, Chuang and Dong, Linhao and Dong, Qianqian and Du, Yujiao and Gao, Kepan and others},
  journal={arXiv preprint arXiv:2407.04675},
  year={2024}
}

@article{gpt-4o,
  title={Gpt-4o system card},
  author={Hurst, Aaron and Lerer, Adam and Goucher, Adam P and Perelman, Adam and Ramesh, Aditya and Clark, Aidan and Ostrow, AJ and Welihinda, Akila and Hayes, Alan and Radford, Alec and others},
  journal={arXiv preprint arXiv:2410.21276},
  year={2024}
}

@article{park2019specaugment,
  title={Specaugment: A simple data augmentation method for automatic speech recognition},
  author={Park, Daniel S and Chan, William and Zhang, Yu and Chiu, Chung-Cheng and Zoph, Barret and Cubuk, Ekin D and Le, Quoc V},
  journal={arXiv preprint arXiv:1904.08779},
  year={2019}
}

@article{mini-omni,
  title={Mini-omni: Language models can hear, talk while thinking in streaming},
  author={Xie, Zhifei and Wu, Changqiao},
  journal={arXiv preprint arXiv:2408.16725},
  year={2024}
}

@inproceedings{LTU,
  title={Listen, think, and understand},
  author={Gong, Yuan and Luo, Hongyin and Liu, Alexander and Karlinsky, Leonid and Glass, James R},
  booktitle={International Conference on Learning Representations},
  volume={2024},
  pages={18516--18545},
  year={2024}
}

@inproceedings{wav2vec,
  title={Fine-tuning wav2vec2 for speaker recognition},
  author={Vaessen, Nik and Van Leeuwen, David A},
  booktitle={ICASSP 2022-2022 IEEE International Conference on Acoustics, Speech and Signal Processing (ICASSP)},
  pages={7967--7971},
  year={2022},
  organization={IEEE}
}

@article{hubert,
  title={Hubert: Self-supervised speech representation learning by masked prediction of hidden units},
  author={Hsu, Wei-Ning and Bolte, Benjamin and Tsai, Yao-Hung Hubert and Lakhotia, Kushal and Salakhutdinov, Ruslan and Mohamed, Abdelrahman},
  journal={IEEE/ACM transactions on audio, speech, and language processing},
  volume={29},
  pages={3451--3460},
  year={2021},
  publisher={IEEE}
}

@article{conformer,
  title={Conformer: Convolution-augmented transformer for speech recognition},
  author={Gulati, Anmol and Qin, James and Chiu, Chung-Cheng and Parmar, Niki and Zhang, Yu and Yu, Jiahui and Han, Wei and Wang, Shibo and Zhang, Zhengdong and Wu, Yonghui and others},
  journal={arXiv preprint arXiv:2005.08100},
  year={2020}
}

@article{pengi,
  title={Pengi: An audio language model for audio tasks},
  author={Deshmukh, Soham and Elizalde, Benjamin and Singh, Rita and Wang, Huaming},
  journal={Advances in Neural Information Processing Systems},
  volume={36},
  pages={18090--18108},
  year={2023}
}

@article{audiopalm,
  title={Audiopalm: A large language model that can speak and listen},
  author={Rubenstein, Paul K and Asawaroengchai, Chulayuth and Nguyen, Duc Dung and Bapna, Ankur and Borsos, Zal{\'a}n and Quitry, F{\'e}lix de Chaumont and Chen, Peter and Badawy, Dalia El and Han, Wei and Kharitonov, Eugene and others},
  journal={arXiv preprint arXiv:2306.12925},
  year={2023}
}

@inproceedings{wavllm,
  title={Wavllm: Towards robust and adaptive speech large language model},
  author={Hu, Shujie and Zhou, Long and Liu, Shujie and Chen, Sanyuan and Meng, Lingwei and Hao, Hongkun and Pan, Jing and Liu, Xunying and Li, Jinyu and Sivasankaran, Sunit and others},
  booktitle={Findings of the Association for Computational Linguistics: EMNLP 2024},
  pages={4552--4572},
  year={2024}
}

@article{audioflamingo,
  title={Audio flamingo: A novel audio language model with few-shot learning and dialogue abilities},
  author={Kong, Zhifeng and Goel, Arushi and Badlani, Rohan and Ping, Wei and Valle, Rafael and Catanzaro, Bryan},
  journal={arXiv preprint arXiv:2402.01831},
  year={2024}
}

@inproceedings{switchboard,
  title={SWITCHBOARD: Telephone speech corpus for research and development},
  author={Godfrey, John J and Holliman, Edward C and McDaniel, Jane},
  booktitle={[Proceedings] ICASSP-92: 1992 IEEE International Conference on Acoustics, Speech, and Signal Processing},
  volume={1},
  pages={517--520},
  year={1992},
  organization={IEEE}
}

@inproceedings{rubostspeechbench,
  title={Speech robust bench: A robustness benchmark for speech recognition},
  author={Shah, Muhammad and Solans Noguero, David and Heikkil{\"a}, Mikko and Raj, Bhiksha and Kourtellis, Nicolas},
  booktitle={International Conference on Learning Representations},
  volume={2025},
  pages={38625--38651},
  year={2025}
}

@article{huang2025rebellion,
  title={Rebellion: Noise-Robust Reasoning Training for Audio Reasoning Models},
  author={Huang, Tiansheng and Shejwalkar, Virat and Chang, Oscar and Nasr, Milad and Liu, Ling},
  journal={arXiv preprint arXiv:2511.09682},
  year={2025}
}

@inproceedings{ami,
  title={The AMI meeting corpus},
  author={Kraaij, Wessel and Hain, Thomas and Lincoln, Mike and Post, Wilfried},
  booktitle={Proc. International Conference on Methods and Techniques in Behavioral Research},
  pages={1--4},
  year={2005}
}

@inproceedings{audio-reasoner,
  title={Audio-reasoner: Improving reasoning capability in large audio language models},
  author={Zhifei, Xie and Lin, Mingbao and Liu, Zihang and Wu, Pengcheng and Yan, Shuicheng and Miao, Chunyan},
  booktitle={Proceedings of the 2025 Conference on Empirical Methods in Natural Language Processing},
  pages={23840--23862},
  year={2025}
}

\clearpage
\appendix

\newpage
\section{Qualitative Case Studies}

We provide representative qualitative examples to illustrate how \textsc{Mega-ASR} changes
the error modes of the baseline under severe acoustic degradation. The examples
cover five common failure patterns: off-audio hallucination, empty-output
collapse, dropout-induced semantic drift, noisy semantic drift, and entity-level
recovery on standard noisy benchmarks. These examples are not intended to
replace quantitative evaluation; instead, they clarify the types of errors that
are reduced by Mega-ASR.

\begin{figure*}[t]
    \centering
    \includegraphics[width=\textwidth]{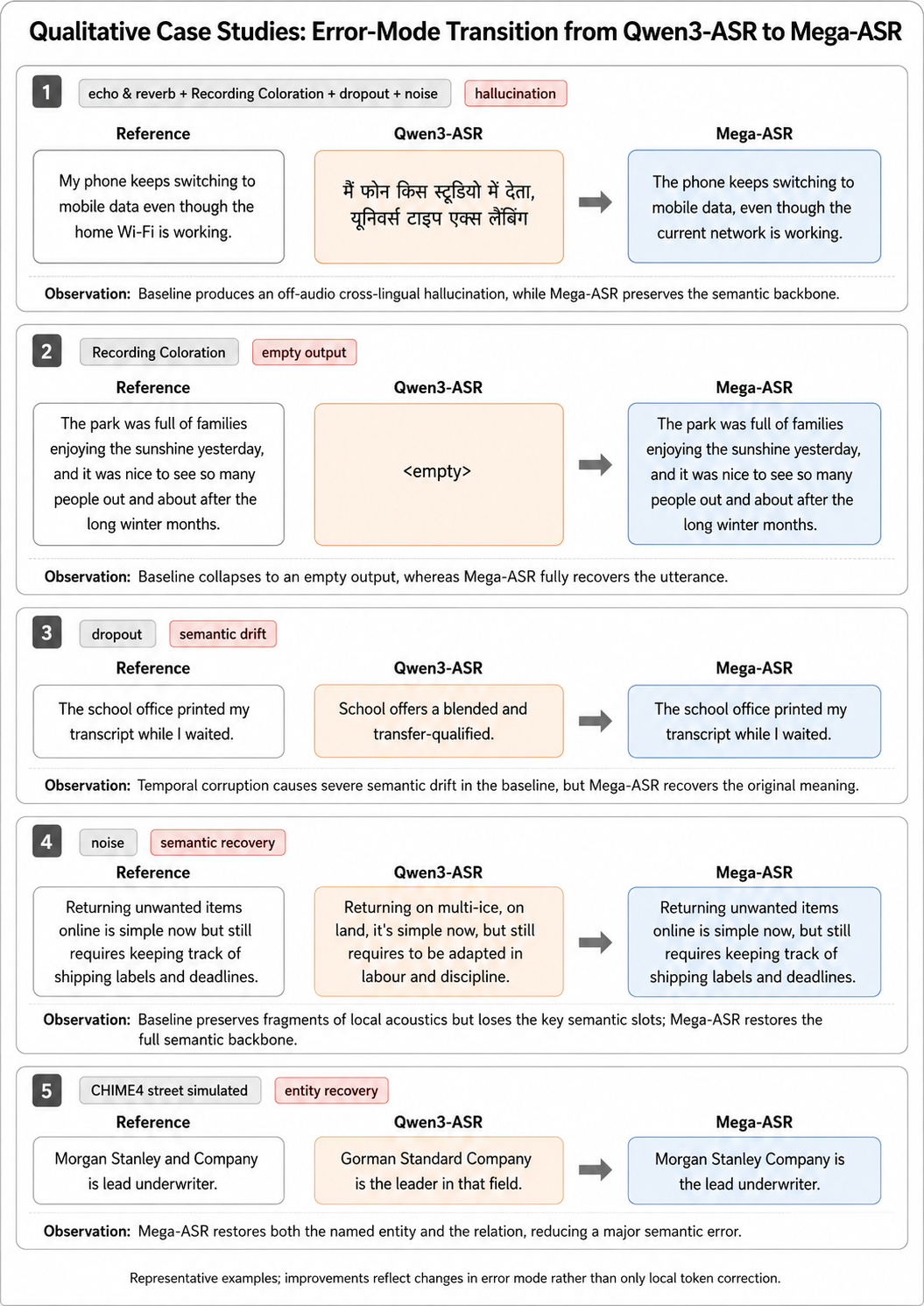}
    \caption{
    Qualitative case studies showing error-mode transitions from Qwen3-ASR to
    Mega-ASR. The examples cover compound acoustic degradation, recording
    coloration, dropout, noise, and CHiME-4 street noise. Compared with the
    baseline, \textsc{Mega-ASR} reduces catastrophic failure modes such as cross-lingual
    hallucination, empty-output collapse, semantic drift, and entity-relation
    errors.
    }
    \label{fig:qualitative_case_studies}
\end{figure*}

\paragraph{Observation.}
Across these examples, the baseline errors are often not local substitutions.
They include cross-lingual hallucination, empty outputs, severe semantic drift,
and missing key entities. \textsc{Mega-ASR} often converts these catastrophic failures
into correct or near-correct transcriptions, preserving the semantic backbone of
the utterance even when minor lexical differences remain.
As shown in Figure~\ref{fig:qualitative_case_studies}, the baseline frequently
fails in ways that are qualitatively different from ordinary word-level
substitutions. In the compound case, it produces a cross-lingual off-audio
hallucination; under recording coloration, it collapses into an empty output;
under dropout and noise, it drifts toward plausible but incorrect semantics; and
on CHiME-4, it changes both the named entity and the relation. \textsc{Mega-ASR} reduces
these catastrophic errors and recovers the semantic backbone of the reference
utterances. These examples support our central observation that severe acoustic
degradation changes the ASR error regime from local recognition errors to
sentence-level semantic failures, and that \textsc{Mega-ASR} mitigates this transition.

\newpage
\section{Additional Robust Benchmark Results}
\label{app:additional_robust_results}

To complement the main results, we provide a focused comparison among
Qwen3-ASR-1.7B, Merged-v2, and the quality-routed 3-LoRA variant on three robust
ASR benchmarks: CHiME-4, NOIZEUS, and VOiCES. These benchmarks cover different
types of adverse acoustic conditions, including real and simulated noisy speech,
controlled additive noise at different SNR levels, and far-field room acoustics.
Table~\ref{tab:robust_three_dataset_summary} summarizes the average WER on each
benchmark, while Tables~\ref{tab:chime4_three_models}--\ref{tab:voices_three_models}
provide detailed subset-level breakdowns.

\begin{table}[t]
\centering
\small
\caption{
Average WER comparison on three robust ASR benchmarks. Lower WER is better.
}
\vspace{-1mm}
\label{tab:robust_three_dataset_summary}
\begin{tabular}{lcccc}
\toprule
\textbf{Model}
& \textbf{CHiME-4}
& \textbf{NOIZEUS}
& \textbf{VOiCES}
& \textbf{Avg.} \\
\midrule
Qwen3-ASR-1.7B
& 5.39 & 9.45 & 8.94 & 7.93 \\

Mega-ASR
& 5.23 & \textbf{7.52} & \textbf{7.35} & \textbf{6.70} \\

Mega-ASR w/ router
& \textbf{5.00} & 7.90 & 7.37 & 6.76 \\
\bottomrule
\end{tabular}
\vspace{-2mm}
\end{table}

Both enhanced variants improve robustness over the Qwen3-ASR-1.7B backbone on
the three-benchmark average. Merged-v2 achieves the best average WER on VOiCES
and NOIZEUS, indicating that always-on robust adaptation is particularly
effective under far-field room acoustics and controlled noisy conditions.
The quality-routed 3-LoRA variant achieves the best average WER on CHiME-4,
suggesting that quality-aware routing is especially useful when the model needs
to balance real and simulated noisy speech while preserving backbone behavior.
Overall, the results show that the proposed robust adaptation improves
recognition accuracy across different adverse acoustic regimes, while the routed
variant provides a practical trade-off between robustness and backbone
preservation.

\begin{table*}[t]
\centering
\caption{
Detailed WER comparison on CHiME-4. Lower WER is better.
}
\vspace{-2mm}
\label{tab:chime4_three_models}
\resizebox{\textwidth}{!}{
\begin{tabular}{lccc}
\toprule
\textbf{Subset}
& \textbf{Qwen3-ASR-1.7B}
& \textbf{Mega-ASR}
& \textbf{Mega-ASR w/ router} \\
\midrule
dt05\_bus\_real & 4.01 & \textbf{3.55} & 3.76 \\
dt05\_bus\_simu & 3.85 & 4.22 & \textbf{3.71} \\
dt05\_caf\_real & 4.00 & \textbf{3.84} & 3.90 \\
dt05\_caf\_simu & 5.92 & 5.94 & \textbf{5.61} \\
dt05\_ped\_real & 3.57 & 3.76 & \textbf{3.50} \\
dt05\_ped\_simu & 4.59 & 4.70 & \textbf{4.22} \\
dt05\_str\_real & \textbf{3.79} & 4.03 & 3.89 \\
dt05\_str\_simu & 5.02 & 5.27 & \textbf{4.62} \\
\midrule
et05\_bus\_real & 6.59 & 6.21 & \textbf{5.94} \\
et05\_bus\_simu & 6.06 & \textbf{5.47} & 5.53 \\
et05\_caf\_real & 5.53 & 5.22 & \textbf{4.89} \\
et05\_caf\_simu & 8.36 & 8.36 & \textbf{7.97} \\
et05\_ped\_real & 4.92 & \textbf{4.53} & 4.71 \\
et05\_ped\_simu & 6.47 & 6.61 & \textbf{6.08} \\
et05\_str\_real & 4.85 & \textbf{4.16} & 4.42 \\
et05\_str\_simu & 8.64 & 7.74 & \textbf{7.19} \\
\midrule
\textbf{Average} & 5.39 & 5.23 & \textbf{5.00} \\
\bottomrule
\end{tabular}
}
\vspace{-2mm}
\end{table*}

\begin{table*}[t]
\centering
\caption{
Detailed WER comparison on NOIZEUS. Lower WER is better.
}
\vspace{-2mm}
\label{tab:noizeus_three_models}
\resizebox{\textwidth}{!}{
\begin{tabular}{lccc}
\toprule
\textbf{Subset}
& \textbf{Qwen3-ASR-1.7B}
& \textbf{Mega-ASR}
& \textbf{Mega-ASR w/ router} \\
\midrule
airport\_0dB & 16.12 & 12.80 & \textbf{12.31} \\
airport\_5dB & 5.37 & \textbf{3.31} & 3.72 \\
airport\_10dB & 2.89 & \textbf{2.07} & 2.89 \\
airport\_15dB & 1.24 & \textbf{0.41} & 1.65 \\
\midrule
babble\_0dB & \textbf{24.79} & 25.20 & 27.43 \\
babble\_5dB & 9.50 & \textbf{5.79} & \textbf{5.79} \\
babble\_10dB & \textbf{2.07} & \textbf{2.07} & 2.48 \\
babble\_15dB & \textbf{1.24} & \textbf{1.24} & \textbf{1.24} \\
\midrule
car\_0dB & 29.34 & 23.90 & \textbf{22.47} \\
car\_5dB & 7.85 & \textbf{5.79} & 6.20 \\
car\_10dB & 2.89 & \textbf{2.07} & 2.89 \\
car\_15dB & 2.07 & \textbf{0.83} & 1.65 \\
\midrule
exhibition\_0dB & 16.12 & \textbf{13.70} & 14.20 \\
exhibition\_5dB & 9.09 & 8.68 & \textbf{5.79} \\
exhibition\_10dB & 3.31 & \textbf{1.65} & 2.89 \\
exhibition\_15dB & 1.65 & \textbf{0.83} & 2.07 \\
\midrule
restaurant\_0dB & 23.14 & 18.10 & \textbf{15.03} \\
restaurant\_5dB & 9.92 & \textbf{7.85} & \textbf{7.85} \\
restaurant\_10dB & 2.89 & 2.89 & \textbf{2.48} \\
restaurant\_15dB & 2.07 & \textbf{0.41} & 1.65 \\
\midrule
station\_0dB & 29.34 & \textbf{21.10} & 23.71 \\
station\_5dB & 6.61 & \textbf{5.37} & 5.79 \\
station\_10dB & 3.31 & \textbf{2.48} & \textbf{2.48} \\
station\_15dB & 1.65 & \textbf{0.83} & 1.65 \\
\midrule
street\_0dB & 28.93 & \textbf{21.90} & 22.47 \\
street\_5dB & 10.74 & \textbf{8.26} & 11.16 \\
street\_10dB & 4.96 & \textbf{4.13} & \textbf{4.13} \\
street\_15dB & 2.48 & \textbf{1.24} & 2.07 \\
\midrule
train\_0dB & 23.97 & 21.70 & \textbf{20.81} \\
train\_5dB & 8.68 & \textbf{7.85} & 9.50 \\
train\_10dB & 4.96 & 4.96 & \textbf{4.13} \\
train\_15dB & 3.31 & \textbf{1.24} & 2.07 \\
\midrule
\textbf{Average} & 9.45 & \textbf{7.52} & 7.90 \\
\bottomrule
\end{tabular}
}
\vspace{-2mm}
\end{table*}
\begin{table*}[t]
\centering
\caption{
Detailed WER comparison on VOiCES. Lower WER is better.
}
\vspace{-2mm}
\label{tab:voices_three_models}
\resizebox{\textwidth}{!}{
\begin{tabular}{lccc}
\toprule
\textbf{Subset}
& \textbf{Qwen3-ASR-1.7B}
& \textbf{Mega-ASR}
& \textbf{Mega-ASR w/ router} \\
\midrule
rm1\_babb\_clo & 2.24 & \textbf{1.94} & 2.16 \\
rm1\_babb\_far & 3.14 & 3.00 & \textbf{2.89} \\
rm1\_musi\_clo & 2.31 & \textbf{2.20} & 2.21 \\
rm1\_musi\_far & 2.77 & 2.72 & \textbf{2.69} \\
rm1\_none\_clo & 2.10 & \textbf{1.95} & 2.12 \\
rm1\_none\_far & \textbf{2.16} & 2.23 & 2.23 \\
rm1\_tele\_clo & 2.44 & \textbf{2.21} & 2.28 \\
rm1\_tele\_far & 3.00 & \textbf{2.66} & 2.76 \\
\midrule
rm2\_babb\_clo & \textbf{2.26} & 2.27 & \textbf{2.26} \\
rm2\_babb\_far & 3.53 & \textbf{3.24} & 3.29 \\
rm2\_musi\_clo & 2.25 & 2.13 & \textbf{2.12} \\
rm2\_musi\_far & 3.14 & \textbf{2.69} & 2.89 \\
rm2\_none\_clo & 2.02 & \textbf{1.96} & 1.98 \\
rm2\_none\_far & 2.41 & \textbf{2.14} & 2.24 \\
rm2\_tele\_clo & 2.28 & \textbf{2.17} & \textbf{2.17} \\
rm2\_tele\_far & 3.08 & \textbf{2.84} & 2.97 \\
\midrule
rm3\_babb\_clo & 7.70 & 6.77 & \textbf{6.50} \\
rm3\_babb\_far & 46.62 & \textbf{36.50} & 37.60 \\
rm3\_musi\_clo & 5.48 & \textbf{4.93} & 5.15 \\
rm3\_musi\_far & 33.83 & \textbf{25.80} & 26.70 \\
rm3\_none\_clo & 3.14 & 2.64 & \textbf{2.62} \\
rm3\_none\_far & 10.35 & \textbf{8.40} & 8.80 \\
rm3\_tele\_clo & 5.96 & \textbf{4.85} & \textbf{4.85} \\
rm3\_tele\_far & 40.40 & 31.15 & \textbf{30.34} \\
\midrule
rm4\_babb\_clo & 2.79 & \textbf{2.56} & 2.71 \\
rm4\_babb\_far & 54.01 & 45.69 & \textbf{43.73} \\
rm4\_musi\_clo & 2.40 & \textbf{1.99} & 2.26 \\
rm4\_musi\_far & 12.43 & 10.54 & \textbf{9.71} \\
rm4\_none\_clo & 2.03 & \textbf{1.91} & 2.08 \\
rm4\_none\_far & 2.69 & \textbf{2.64} & \textbf{2.64} \\
rm4\_tele\_clo & 2.18 & \textbf{2.02} & 2.12 \\
rm4\_tele\_far & 12.95 & \textbf{8.36} & 8.80 \\
\midrule
\textbf{Average} & 8.94 & \textbf{7.35} & 7.37 \\
\bottomrule
\end{tabular}
}
\vspace{-2mm}
\end{table*}

\newpage
\section{Details of  \textsc{Voices-in-the-wild-2M}  Construction}
\label{app:data_construction}

\subsection{Hierarchical Simulation Pipeline}

 \textsc{Voices-in-the-wild-2M}  is constructed through a hierarchical acoustic simulation
pipeline. Rather than directly enumerating complex real-world environments, we
decompose in-the-wild speech degradation into three levels: primitive acoustic
effects, atomic acoustic effects, and compound acoustic scenarios.

At the lowest level, we define eight primitive acoustic effects, each
corresponding to an independent and controllable signal-level transformation:
additive noise, echo delay, reverberation, nonlinear distortion, resampling,
spectral filtering, loudness transformation, and frame-level stutter. These
primitive effects are designed to capture basic physical or device-induced
degradation mechanisms, such as background interference, delayed reflection,
room reverberation, clipping, bandwidth limitation, spectral attenuation, gain
mismatch, and packet loss.

At the intermediate level, we construct seven atomic acoustic effects from these
primitive effects: \textit{noise}, \textit{far-field}, \textit{obstructed},
\textit{echo\&reverb}, \textit{recording}, \textit{electronic distortion}, and
\textit{transmission dropout}. Each atomic effect is not necessarily implemented
by a single primitive effect. Instead, it is instantiated as a physically
motivated composition of one dominant primitive effect and several auxiliary
primitive effects. For example, far-field speech is not only quieter, but also
more reverberant and spectrally attenuated; similarly, low-quality recording may
simultaneously involve bandwidth limitation, gain mismatch, and channel
coloration.

At the highest level, we construct compound acoustic scenarios by composing
multiple atomic acoustic effects. This produces complex acoustic environments
that better match in-the-wild speech, where multiple degradation sources often
co-occur. Importantly, during both atomic-effect construction and compound
scenario construction, we preserve a fixed topological order among primitive
effects. This avoids physically implausible processing chains and ensures that
the same low-level degradation mechanism is applied consistently across
different scenarios. The final pipeline can therefore be summarized as
\[
\text{8 primitive effects}
\rightarrow
\text{7 atomic acoustic effects}
\rightarrow
\text{54 compound acoustic scenarios}.
\]
\subsection{Primitive Acoustic Effects}

\paragraph{Motivation.}
The seven atomic acoustic effects used in the main paper are high-level
descriptions of real-world acoustic phenomena. However, such phenomena are
usually caused by multiple lower-level signal transformations. For example,
speech behind a door may be attenuated, low-pass filtered, and slightly
reverberant; speech transmitted through an unstable communication channel may
contain repeated frames, local dropouts, and bandwidth loss. We therefore first
define a set of primitive acoustic effects, which serve as the basic
signal-level operators for building both atomic and compound scenarios.

Each primitive effect exposes a small number of interpretable parameters. These
parameters control the strength of the degradation and are later tied to the
global severity variable used in dataset synthesis. The primitive effects are
kept modular, allowing them to be composed while preserving a consistent
topological order.

\begin{table*}[t]
\centering
\small
\setlength{\tabcolsep}{4pt}
\renewcommand{\arraystretch}{1.15}
\caption{Eight primitive acoustic effects used as signal-level building blocks in  \textsc{Voices-in-the-wild-2M} .}
\label{tab:primitive_effects}
\begin{tabular}{p{0.16\textwidth} p{0.25\textwidth} p{0.25\textwidth} p{0.24\textwidth}}
\toprule
\textbf{Primitive effect} & \textbf{Main parameters} & \textbf{Simulated degradation} & \textbf{Typical real-world source} \\
\midrule
Additive noise
& noise source, noise category, relative noise level, wet ratio
& Background interference from environmental sounds, human voices, or device noise
& Street, office, vehicle, crowd, household environment \\

Echo delay
& delay time, feedback, mix ratio
& Discrete delayed reflections and repeated copies of speech
& Empty room, corridor, tunnel, large hall \\

Reverberation
& room size, damping, wet level, dry level
& Dense room reflections and long-tail spatial smearing
& Classroom, auditorium, church, meeting room \\

Nonlinear distortion
& drive gain, wet ratio
& Overload, saturation, clipping, and harmonic artifacts
& Low-quality microphone, over-amplified recorder, damaged device \\

Resampling
& target sampling rate, wet ratio, probability gate
& Bandwidth limitation and high-frequency information loss
& Telephone channel, compressed audio, low-bandwidth transmission \\

Spectral filtering
& filter type, cutoff frequency, repeat count, wet ratio
& Frequency attenuation and channel coloration
& Mask, door, wall, glass, narrow-band device \\

Loudness transformation
& target LUFS
& Gain mismatch, distance-induced attenuation, or abnormal recording level
& Distant speaker, quiet recording, over-amplified microphone \\

Frame-level stutter
& frame length, stutter probability, repeat probability, maximum repeats
& Local dropout, repeated frames, and unstable temporal continuity
& Packet loss, unstable streaming, corrupted recording \\
\bottomrule
\end{tabular}
\end{table*}

\paragraph{Additive noise.}
The additive-noise primitive mixes an external noise waveform into the clean
speech signal. The noise source can be selected from a specified noise category
or from a given noise file. If the noise waveform is shorter than the speech
signal, it is tiled and then cropped to match the speech duration. The noise is
RMS-normalized according to a target relative level and then mixed with the
clean speech using a wet ratio. This primitive captures background interference
from environmental sounds, human voices, and device noise.

\paragraph{Echo delay.}
The echo-delay primitive adds delayed copies of the original signal. It is
controlled by the delay time, feedback strength, and dry-wet mix ratio. The
delay time determines the temporal offset between the direct path and the
reflected path, while the feedback parameter controls the strength and number
of repeated reflections. This primitive mainly simulates sparse and perceptible
echoes in highly reflective spaces.

\paragraph{Reverberation.}
The reverberation primitive simulates dense room reflections. It is controlled
by room size, damping, wet level, and dry level. Larger room size and higher wet
level produce stronger spatial smearing, while damping controls the decay of
high-frequency components in the reverberant tail. Unlike echo delay, which
models discrete delayed repetitions, reverberation models dense and continuous
reflection patterns.

\paragraph{Nonlinear distortion.}
The nonlinear-distortion primitive applies overdrive to the waveform and
produces saturation or clipping artifacts. The drive gain controls the strength
of the distortion: small values introduce mild coloration, whereas large values
produce clear overload artifacts and additional harmonics. After distortion, the
output is clipped to the valid amplitude range, further simulating harsh device
overload.

\paragraph{Resampling.}
The resampling primitive first downsamples the waveform to a lower target
sampling rate and then upsamples it back to the original sampling rate. This
removes high-frequency details and introduces bandwidth limitation while keeping
the final sampling rate compatible with the rest of the pipeline. A probability
gate is used so that resampling can be applied stochastically when constructing
mixed scenarios.

\paragraph{Spectral filtering.}
The spectral-filtering primitive applies either a low-pass or high-pass filter
with a specified cutoff frequency. The filter can be repeatedly applied to
increase the strength of spectral attenuation. Low-pass filtering removes
high-frequency details and is useful for muffled or occluded speech, while
high-pass filtering removes low-frequency energy and simulates thin channel
responses or device coloration.

\paragraph{Loudness transformation.}
The loudness primitive adjusts the signal to a target LUFS value. Unlike simple
amplitude scaling, LUFS normalization provides a perceptually meaningful measure
of loudness. This primitive is used to simulate distance-induced attenuation,
microphone gain mismatch, quiet speech, and over-amplified recordings. When the
loudness of extremely short or silent audio cannot be estimated reliably, the
original signal is kept unchanged.

\paragraph{Frame-level stutter.}
The frame-level stutter primitive partitions the waveform into short frames and
randomly triggers local replacement events. Once triggered, several consecutive
frames are either replaced by the previous frame or replaced by silence. The
total audio length is kept unchanged, which allows the resulting audio to remain
aligned with the original transcript while still containing local temporal
discontinuities. This primitive simulates packet loss, unstable streaming,
frame repetition, and local dropout.

\subsection{Construction of Seven Atomic Acoustic Effects}

Based on the eight primitive acoustic effects, we further construct seven
atomic acoustic effects that correspond to common in-the-wild acoustic
conditions: \textit{noise}, \textit{far-field}, \textit{obstructed},
\textit{echo\&reverb}, \textit{recording coloration},
\textit{electronic distortion}, and \textit{transmission dropout}. Each atomic
effect is implemented as an ordered chain of primitive effects. The primitive
chain is designed to make one degradation mechanism dominant while retaining
secondary artifacts that naturally co-occur in the corresponding real-world
condition.

Table~\ref{tab:atomic_effect_chains} first provides a structural overview of the
seven atomic acoustic effects. It reports the ordered primitive-effect chain,
the dominant degradation mechanism, and the representative real-world condition
for each atomic effect. The listed order follows the corresponding scene
configuration rather than a manually imposed global order.

To make the simulation fully reproducible, Table~\ref{tab:atomic_effect_parameters}
further summarizes the key parameters used in each scene configuration. We group
parameters by primitive effect and distinguish randomly sampled ranges from
fixed values. Parameters marked with ``core'' are the primary severity-controlling
parameters used to modulate the difficulty of the corresponding atomic effect.

\begin{table*}[t]
\centering
\small
\setlength{\tabcolsep}{4pt}
\renewcommand{\arraystretch}{1.15}
\caption{Construction of seven atomic acoustic effects from primitive acoustic effects.}
\label{tab:atomic_effect_chains}
\begin{tabular}{p{0.16\textwidth} p{0.34\textwidth} p{0.20\textwidth} p{0.22\textwidth}}
\toprule
\textbf{Atomic effect} & \textbf{Primitive-effect chain} & \textbf{Dominant degradation} & \textbf{Representative condition} \\
\midrule
Noise
& \texttt{add\_noise} $\rightarrow$ \texttt{change\_volume}
& Low signal-to-noise ratio
& Street, cafe, vehicle, crowd \\

Far-field
& \texttt{add\_reverb} $\rightarrow$ \texttt{apply\_filter} $\rightarrow$ \texttt{change\_volume}
& Distance-induced reverberation and attenuation
& Speaking to a distant microphone \\

Obstructed
& \texttt{apply\_filter} $\rightarrow$ \texttt{add\_reverb} $\rightarrow$ \texttt{change\_volume}
& Occlusion-induced spectral loss
& Speech behind a wall, door, or mask \\

Echo\&reverb
& \texttt{add\_reverb} $\rightarrow$ \texttt{apply\_filter} $\rightarrow$ \texttt{add\_echo} $\rightarrow$ \texttt{change\_volume}
& Strong reflections and delayed echoes
& Gymnasium, garage, large hall \\

Recording Coloration
& \texttt{add\_resample} $\rightarrow$ \texttt{add\_noise} $\rightarrow$ \texttt{apply\_filter} $\rightarrow$ \texttt{apply\_filter} $\rightarrow$ \texttt{change\_volume}
& Playback-recording channel degradation
& Phone playback recorded by another device \\

Electronic distortion
& \texttt{add\_distortion} $\rightarrow$ \texttt{apply\_filter} $\rightarrow$ \texttt{change\_volume\_distortion}
& Clipping and nonlinear overload
& Close-talking with excessive recording gain \\

Transmission dropout
& \texttt{add\_stutter\_replace} $\rightarrow$ \texttt{change\_volume}
& Local temporal discontinuity
& VoIP packet loss, unstable Bluetooth or streaming \\
\bottomrule
\end{tabular}
\end{table*}

\begin{table*}[t]
\centering
\scriptsize
\setlength{\tabcolsep}{3.5pt}
\renewcommand{\arraystretch}{1.18}
\caption{Parameterization of the seven atomic acoustic effects. Randomly sampled parameters are shown as ranges, while fixed parameters are listed separately. Core parameters are the main severity-controlling variables in the corresponding scene configuration.}
\label{tab:atomic_effect_parameters}
\begin{tabular}{p{0.14\textwidth} p{0.18\textwidth} p{0.39\textwidth} p{0.21\textwidth}}
\toprule
\textbf{Atomic effect} & \textbf{Primitive effect} & \textbf{Sampled severity parameters} & \textbf{Fixed parameters} \\
\midrule

\multirow{2}{*}{Noise}
& \texttt{add\_noise}
& \texttt{noise\_db}$\in[-5,10]$ \textbf{(core)}
& \texttt{noise\_category=filtered\_wavs}, \texttt{wet=1.0} \\
& \texttt{change\_volume}
& --
& \texttt{target\_lufs=-23.0} \\

\midrule
\multirow{3}{*}{Far-field}
& \texttt{add\_reverb}
& \texttt{room\_size}$\in[0.4,0.6]$ \textbf{(core)}, \texttt{damping}$\in[0.6,0.8]$, \texttt{wet\_level}$\in[0.4,0.5]$
& \texttt{dry\_level=0.5} \\
& \texttt{apply\_filter}
& \texttt{cutoff\_hz}$\in[3500,4500]$ \textbf{(core)}
& \texttt{filter\_type=lowpass}, \texttt{repeat=3}, \texttt{wet=1.0} \\
& \texttt{change\_volume}
& \texttt{target\_lufs}$\in[-38,-27]$ \textbf{(core)}
& -- \\

\midrule
\multirow{3}{*}{Obstructed}
& \texttt{apply\_filter}
& \texttt{cutoff\_hz}$\in[1500,2000]$ \textbf{(core)}, \texttt{repeat}$\in\{2,3,4\}$
& \texttt{filter\_type=lowpass}, \texttt{wet=0.9} \\
& \texttt{add\_reverb}
& \texttt{wet\_level}$\in[0.5,0.7]$
& \texttt{room\_size=0.4}, \texttt{damping=0.9}, \texttt{dry\_level=0.4} \\
& \texttt{change\_volume}
& \texttt{target\_lufs}$\in[-25,-15]$ \textbf{(core)}
& -- \\

\midrule
\multirow{4}{*}{Echo\&reverb}
& \texttt{add\_reverb}
& \texttt{room\_size}$\in[0.8,0.95]$ \textbf{(core)}, \texttt{wet\_level}$\in[0.6,0.8]$
& \texttt{damping=0.5}, \texttt{dry\_level=0.4} \\
& \texttt{apply\_filter}
& \texttt{cutoff\_hz}$\in[100,300]$
& \texttt{filter\_type=highpass}, \texttt{repeat=1}, \texttt{wet=1.0} \\
& \texttt{add\_echo}
& \texttt{delay\_seconds}$\in[0.1,0.3]$ \textbf{(core)}, \texttt{feedback}$\in[0.3,0.5]$, \texttt{mix}$\in[0.2,0.3]$
& -- \\
& \texttt{change\_volume}
& \texttt{target\_lufs}$\in[-30,-23]$ \textbf{(core)}
& -- \\

\midrule
\multirow{5}{*}{Recording}
& \texttt{add\_resample}
& \texttt{prob}$\in[0,1]$ \textbf{(core)}
& \texttt{target\_sr=8000}, \texttt{wet=1.0}, \texttt{threshold=0.4} \\
& \texttt{add\_noise}
& \texttt{noise\_db}$\in[-5,10]$ \textbf{(core)}
& \texttt{use\_white\_noise=True}, \texttt{wet=1.0} \\
& \texttt{apply\_filter}
& \texttt{cutoff\_hz}$\in[400,600]$ \textbf{(core)}, \texttt{repeat}$\in\{4,5,6\}$
& \texttt{filter\_type=highpass}, \texttt{wet=1.0} \\
& \texttt{apply\_filter}
& \texttt{cutoff\_hz}$\in[3500,4500]$ \textbf{(core)}, \texttt{repeat}$\in\{4,5,6\}$
& \texttt{filter\_type=lowpass}, \texttt{wet=1.0} \\
& \texttt{change\_volume}
& --
& \texttt{target\_lufs=-23.0} \\

\midrule
\multirow{3}{*}{Electronic distortion}
& \texttt{add\_distortion}
& \texttt{drive\_db}$\in[20,60]$ \textbf{(core)}
& \texttt{wet=1.0} \\
& \texttt{apply\_filter}
& \texttt{cutoff\_hz}$\in[2800,6000]$
& \texttt{filter\_type=lowpass}, \texttt{repeat=1}, \texttt{wet=1.0} \\
& \texttt{change\_volume\_distortion}
& \texttt{target\_lufs}$\in[-38,-27]$ \textbf{(core)}
& -- \\

\midrule
\multirow{2}{*}{Transmission dropout}
& \texttt{add\_stutter\_replace}
& \texttt{stutter\_prob}$\in[0.05,0.3]$ \textbf{(core)}, \texttt{max\_repeats}$\in\{2,3,4\}$
& \texttt{repeat\_prob=0.7}, \texttt{frame\_ms=20} \\
& \texttt{change\_volume}
& --
& \texttt{target\_lufs=-23.0} \\

\bottomrule
\end{tabular}
\end{table*}

\paragraph{Noise.}
The noise atomic effect is designed to isolate low-SNR recognition difficulty.
It therefore uses additive noise as the dominant primitive effect and avoids
introducing strong reverberation or filtering artifacts. The noise source is
sampled from the prepared noise pool, and its relative level is varied to
produce different SNR regimes. A final loudness normalization step keeps the
overall output level comparable across samples, ensuring that the primary
challenge comes from masking rather than from abnormal global volume. This
design matches common noisy environments such as streets, cafes, vehicles, and
crowded rooms.

\paragraph{Far-field.}
The far-field atomic effect models speech captured by a distant microphone.
Its primitive chain first introduces room reverberation, then applies low-pass
filtering to mimic high-frequency attenuation, and finally reduces the loudness
to simulate distance-induced energy decay. This combination reflects the main
acoustic properties of far-field speech: stronger room reflections, weaker
direct-path energy, and mild spectral attenuation. The resulting samples target
scenarios such as speaking to a smart speaker from across a room.

\paragraph{Obstructed.}
The barrier atomic effect simulates speech transmitted through an obstacle,
such as a wall, door, glass, or mask. The dominant operation is low-pass
filtering, which removes high-frequency components that are difficult to
transmit through physical barriers. The filter can be repeatedly applied to
represent thicker or more absorptive obstacles. We then add reverberation with
a relatively high wet component, reflecting the fact that the listener often
receives a mixture of attenuated direct speech and room-reflected sound from
the other side of the barrier. Finally, the signal is attenuated through
loudness transformation. This makes the generated speech muffled, weaker, and
less spectrally detailed.

\paragraph{Echo\&reverb.}
The strong-echo atomic effect targets highly reflective environments. It
combines dense reverberation with a separate echo-delay primitive. The
reverberation component produces a long reflection tail, while the echo-delay
component introduces perceptible delayed copies of the speech signal. A mild
high-pass filter is additionally applied to control low-frequency muddiness,
and the final loudness transformation keeps the generated samples within a
reasonable intensity range. This construction is suitable for large empty
spaces, underground garages, gymnasiums, and other environments where both
reverberant smearing and discrete echoes are present.

\paragraph{Recording coloration.}
The recording or acoustic-crosstalk atomic effect simulates a playback-recording
loop, such as playing speech from one phone and recording it with another
device. This chain first applies resampling to model bandwidth limitation, then
adds white or device-like noise, and subsequently applies both high-pass and
low-pass filtering. The high-pass filter removes low-frequency energy and makes
the signal thinner, while the low-pass filter limits the upper bandwidth of the
playback-recording channel. A final loudness normalization step standardizes
the output level. Together, these operations produce speech that is narrower in
frequency response, noisier, and more blurred than the original recording.

\paragraph{Electronic distortion.}
The electronic-distortion atomic effect focuses on nonlinear device overload.
It uses distortion as the dominant primitive effect, where larger drive values
produce stronger saturation and clipping. A subsequent low-pass filter mimics
the limited response of low-quality microphone hardware under large input
dynamics, while the final loudness adjustment controls the output level. Unlike
far-field or strong-echo conditions, this scene intentionally avoids adding
reverberation or background noise, so that the dominant challenge remains
waveform-level clipping and harmonic distortion rather than room acoustics or
SNR degradation.

\paragraph{Transmission dropout.}
The transmission-dropout atomic effect models temporal corruption rather than
spectral coloration. It uses frame-level stutter as the dominant primitive
effect: short frames are randomly replaced by previous frames or by silence,
creating local repetitions and dropouts while keeping the total audio length
unchanged. A final loudness normalization step keeps the recording level
standard. We intentionally avoid additional filtering because network or
Bluetooth instability does not necessarily make the speech spectrally muffled.
This effect therefore isolates temporal discontinuity caused by VoIP packet
loss, unstable wireless links, or corrupted streaming.

These seven atomic acoustic effects form the basic scenario vocabulary of
 \textsc{Voices-in-the-wild-2M} . Each one emphasizes a distinct degradation mechanism,
while still including the secondary primitive effects required to make the
simulation realistic. In the next subsection, we use these atomic effects as
building blocks for constructing compound acoustic scenarios.

\subsection{Construction of Compound Acoustic Scenarios}

The seven atomic acoustic effects above serve as the basic scenario vocabulary
for constructing compound acoustic scenarios. However, not all atomic effects
play the same role in real-world acoustic environments. We therefore divide
them into two groups: \textit{scene-defining anchor effects} and
\textit{portable modifier effects}.

The scene-defining anchor effects include \textit{far-field},
\textit{Echo\&reverb}, and \textit{obstructed}. These effects usually determine the
dominant acoustic geometry of a recording condition. For example, far-field
speech is primarily characterized by distance-induced attenuation and
reverberation; strong echo corresponds to highly reflective spaces with delayed
reflections; and barrier speech is dominated by occlusion-induced spectral
attenuation. Since these effects describe mutually distinctive propagation
conditions, we do not directly combine multiple anchor effects within the same
scenario.

The portable modifier effects include \textit{recording coloration},
\textit{electronic distortion}, \textit{noise}, and \textit{transmission
dropout}. These effects are more flexible and can be attached to different
anchor conditions. They correspond to playback-recording artifacts, device
overload, background interference, and unstable transmission, respectively.
Such factors commonly co-occur with different acoustic geometries in real
deployments. For example, far-field speech may also be noisy and distorted, and
barrier speech may additionally suffer from recording-channel degradation.

\paragraph{Scenario enumeration.}
Following this anchor--modifier decomposition, we enumerate 54 acoustic
scenario categories in total. The enumeration consists of four groups.

First, we include the seven single-effect scenarios, corresponding to the seven
atomic acoustic effects themselves. Second, we construct 18 two-effect
scenarios, including all anchor--modifier pairs and all modifier--modifier
pairs. Third, we construct 13 three-effect scenarios. These include
anchor-prefixed combinations with two selected modifiers, as well as all
three-way combinations among the four modifier effects. Finally, we construct
16 higher-order scenarios, including anchor-prefixed combinations with three or
four modifiers and the modifier-only four-way combination.

\begin{table*}[t]
\centering
\small
\setlength{\tabcolsep}{5pt}
\renewcommand{\arraystretch}{1.15}
\caption{Enumeration of the 54 acoustic scenario categories in  \textsc{Voices-in-the-wild-2M} . Anchor effects are \textit{far-field}, \textit{echo\&reverb}, and \textit{obstructed}; modifier effects are \textit{recording coloration}, \textit{electronic distortion}, \textit{noise}, and \textit{transmission dropout}.}
\label{tab:compound_scenario_count}
\begin{tabular}{p{0.22\textwidth} p{0.48\textwidth} p{0.16\textwidth}}
\toprule
\textbf{Group} & \textbf{Construction rule} & \textbf{Number} \\
\midrule
Single-effect scenarios
& Seven atomic acoustic effects
& $7$ \\

Two-effect scenarios
& Anchor--modifier pairs: $3 \times 4$; modifier--modifier pairs: $\binom{4}{2}$
& $12 + 6 = 18$ \\

Three-effect scenarios
& Anchor with two selected modifiers: $3 \times 3$; modifier-only triples: $\binom{4}{3}$
& $9 + 4 = 13$ \\

Higher-order scenarios
& Anchor with three modifiers: $3 \times \binom{4}{3}$; modifier-only four-way combination: $1$; anchor with all four modifiers: $3$
& $12 + 1 + 3 = 16$ \\
\midrule
\textbf{Total}
& --
& $\textbf{54}$ \\
\bottomrule
\end{tabular}
\end{table*}

\paragraph{Anchor--modifier composition.}
The anchor effects define the main acoustic environment, while the modifier
effects introduce additional degradations that are portable across environments.
This design avoids unrealistic combinations among mutually distinctive
propagation geometries. For example, \textit{far-field}, \textit{echo\&reverb},
and \textit{obstructed} each describe a different dominant acoustic path, and
therefore are not directly combined with each other. In contrast, modifiers such
as \textit{noise}, \textit{distortion}, \textit{recording coloration},
and \textit{transmission dropout} can naturally co-occur with many acoustic
geometries.

For two-effect scenarios, we include all anchor--modifier pairs and all
modifier--modifier pairs. For three-effect scenarios, we include two types of
compositions: an anchor effect combined with two modifiers, and modifier-only
triples. For higher-order scenarios, we further include anchor-prefixed
combinations with three modifiers, the modifier-only four-way combination, and
anchor-prefixed combinations with all four modifiers. This enumeration yields a
balanced set of atomic, moderate-composition, and high-composition acoustic
conditions.

\paragraph{Effect-chain maintenance.}
Each compound scenario is represented by a list of atomic effects. To generate
the final signal-processing chain, we merge the ordered primitive-effect chains
of the selected atomic effects. The merge procedure preserves the within-scene
order of each atomic effect and removes cross-scene duplicate primitive effects,
except for additive noise. This exception is used because real environments may
contain multiple independent noise sources. The resulting merged chain is then
parameterized and applied sequentially to the waveform.

\begin{algorithm}[t]
\caption{Effect-chain maintenance for compound acoustic scenarios}
\label{alg:effect_chain_maintenance}
\begin{algorithmic}[1]
\Require Atomic scene configurations $\mathcal{C}$; selected atomic effects $S=[s_1,\ldots,s_m]$
\Require Duplicate-allowed primitive set $\mathcal{D}=\{\texttt{add\_noise}\}$
\State Initialize merged chain $M \leftarrow [\,]$
\State Initialize previously seen primitive set $V \leftarrow \emptyset$
\For{$s_i$ in $S$}
    \State Load ordered primitive-effect chain $E_i$ from $\mathcal{C}[s_i]$
    \State Initialize current-scene primitive set $U \leftarrow \emptyset$
    \For{primitive effect $e$ in $E_i$}
        \If{$e.\mathrm{name} \in \mathcal{D}$}
            \State Append $e$ to $M$
        \ElsIf{$e.\mathrm{name} \notin V$}
            \State Append $e$ to $M$
            \State Add $e.\mathrm{name}$ to $U$
        \Else
            \State Skip $e$ as a cross-scene duplicate
        \EndIf
    \EndFor
    \State $V \leftarrow V \cup U$
\EndFor
\State \Return merged primitive-effect chain $M$
\end{algorithmic}
\end{algorithm}

This merge strategy is important for preserving atomic-effect definitions. For
example, the recording coloration effect intentionally contains two
filtering operations, one high-pass and one low-pass, to narrow the frequency
response. Such within-scene repeated operators are preserved, while duplicate
operators introduced by different atomic effects are removed unless explicitly
allowed.

\subsection{Severity Sampling and Difficulty Calibration}

To make the simulated data both diverse and controllable, we associate each
generated sample with a global severity variable $m \in [0,1]$. Rather than
sampling every effect parameter independently, we first sample a latent variable
$x \sim \mathcal{U}(0,1)$ and then map it to the final severity value $m$ using
a predefined difficulty mapping function. The resulting $m$ is shared across the
primitive effects in the same sample, which ensures that the degradation level
remains globally coherent instead of varying arbitrarily across different
effects.

Formally, for each generated sample we first draw
\[
x \sim \mathcal{U}(0,1),
\]
and compute
\[
m = f(x),
\]
where $f(\cdot)$ is one of four candidate mapping functions. We consider the
following mappings:

\begin{equation}
m_{\text{linear}}(x) = x,
\end{equation}

\begin{equation}
m_{\text{sqrt-fwd}}(x) = \sqrt{x},
\end{equation}

\begin{equation}
m_{\text{sqrt-bwd}}(x) = x^2,
\end{equation}

\begin{equation}
m_{\text{gaussian-mid}}(x)
=
\mathrm{clip}\!\left(
\Phi^{-1}(0.05 + 0.9x; \mu=0.5,\sigma),
\,0,\,1
\right),
\end{equation}
where $\Phi^{-1}(\cdot;\mu,\sigma)$ denotes the inverse CDF of a Gaussian
distribution with mean $\mu=0.5$, and $\sigma$ is set such that the central
region is emphasized while the two extremes are compressed. In practice, this
mapping increases the density of medium-difficulty samples and avoids
over-sampling the easiest and hardest regimes.

The four mappings differ in how they distribute probability mass over the final
severity variable $m$. The linear mapping preserves the original uniform
sampling and therefore distributes difficulty evenly over the full range. The
sqrt-forward mapping allocates more samples to the hard regime by increasing
$m$ rapidly at small $x$. In contrast, the sqrt-backward mapping biases the
sampling toward easier samples, since $m$ grows more slowly at small $x$. The
gaussian-mid mapping concentrates more samples around intermediate difficulty
levels and suppresses both extremes.

\begin{figure}[!h]
    \centering
    \includegraphics[width=0.72\linewidth]{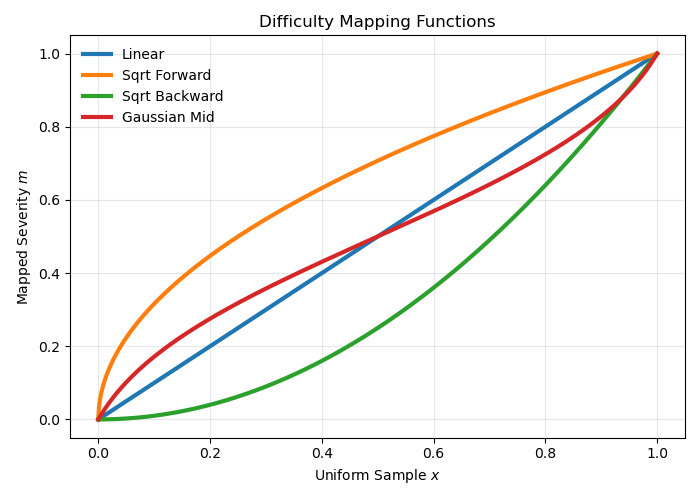}
    \caption{Difficulty mapping functions used to transform a uniform sample
    $x \in [0,1]$ into the final global severity variable $m \in [0,1]$. Linear
    preserves a uniform severity profile; Sqrt Forward emphasizes hard samples;
    Sqrt Backward emphasizes easy samples; and Gaussian Mid concentrates samples
    around the medium-difficulty region.}
    \label{fig:difficulty_mapping_functions}
\end{figure}

Figure~\ref{fig:difficulty_mapping_functions} visualizes the four mapping
functions. Although they all map the same uniform variable $x$ into the common
severity range $[0,1]$, they induce substantially different difficulty profiles
for the generated dataset.

After obtaining the global severity value $m$, we use it to instantiate the
random parameters in each primitive effect. Each random parameter is defined by
a range and a monotonicity flag indicating whether smaller values are easier or
harder. For a parameter with range $[a,b]$, the sampled value is computed as
\[
\theta =
\begin{cases}
a + (b-a)m, & \text{if larger values correspond to harder samples},\\[3pt]
b - (b-a)m, & \text{if smaller values correspond to harder samples}.
\end{cases}
\]
Integer-valued parameters are rounded to the nearest valid integer after this
mapping. For categorical parameters, the same severity value $m$ is used to
select an option index from an ordered candidate list.

This design gives us a unified severity interface across heterogeneous effects.
For example, a larger $m$ may correspond to lower cutoff frequency in a
low-pass filter, larger distortion drive, stronger reverberation, lower target
loudness, or higher stutter probability, depending on the semantics of the
parameter. Importantly, because all parameters in the same sample share the same
global severity variable, the resulting degradation remains internally
consistent: a hard sample tends to be hard across all of its active primitive
effects, while an easy sample remains globally mild.

We empirically compare the four candidate mappings by generating probe sets
under each mapping and evaluating the resulting training utility on real noisy
speech. The goal is not only to increase nominal difficulty, but to obtain a
severity profile that yields the best downstream robustness after supervised
fine-tuning. Among the four candidates, the linear mapping provides the most
balanced coverage of easy, medium, and hard samples, and leads to the best
overall robustness in our pilot experiments. We therefore adopt the linear
mapping as the default severity profile in  \textsc{Voices-in-the-wild-2M} .

Intuitively, the sqrt-forward mapping over-emphasizes hard samples, which may
reduce learnability in the early stages of training; the sqrt-backward mapping
places too much mass on easy samples and therefore under-exposes the model to
challenging conditions; and the gaussian-mid mapping improves coverage of
medium-difficulty samples but under-represents the two ends of the spectrum.
The linear mapping strikes the best balance between coverage, learnability, and
difficulty diversity.

\paragraph{Implementation detail: global severity sharing.}
In our implementation, we use a shared global severity value for all primitive
effects within the same sample. Concretely, a single $x$ is first sampled and
mapped into a single $m$, and this $m$ is then reused when resolving the random
parameters of all active primitive effects in the corresponding effect chain.
This mechanism avoids internally inconsistent mixtures such as a sample with
extremely strong reverberation but almost negligible noise, or severe dropout
combined with otherwise near-clean recording quality. In this way, the sampled
difficulty more faithfully reflects a coherent acoustic condition rather than an
arbitrary mixture of independently sampled parameter strengths.

\newpage
\section{Router Implementation and Training Details}

\subsection{Motivation}

Mega-ASR is optimized for acoustically degraded speech, but always using the
robust weights is not necessarily optimal for all inputs. In particular, the
original Qwen3-ASR backbone retains strong clean-domain behavior and can better
preserve complementary capabilities such as clean-speech recognition, hotword
recognition, and streaming-style inference. We therefore introduce a lightweight
environment-aware router that predicts whether an input utterance should be
processed by the original backbone or by the robust LoRA-enhanced Mega-ASR
weights.

The router is used only for model selection. It does not generate transcripts
and does not modify the ASR decoding process. Given an input audio clip, the
router outputs a binary decision: clean inputs are routed to the base Qwen3-ASR
model, while degraded inputs are routed to the \textsc{Mega-ASR} LoRA branch. This makes
Mega-ASR a plug-and-play robustness module rather than a full replacement of the
original ASR system.

\subsection{Router Model Architecture}

The router is implemented as a lightweight audio-quality classifier. It takes
log-Mel acoustic features as input and predicts a binary label indicating
whether the input is clean or degraded. We use a single-layer Transformer
architecture to minimize routing overhead.

The model first extracts 80-dimensional log-Mel features from the waveform. A
lightweight convolutional frontend maps the Mel features to a hidden dimension
and performs temporal downsampling. The downsampled sequence is then augmented
with sinusoidal positional encoding and passed through a single Transformer
encoder layer. Finally, an attention-pooling module aggregates the frame-level
representations into an utterance-level embedding, followed by a linear binary
classification head.

\begin{table}[t]
\centering
\small
\caption{Architecture of the environment-aware router.}
\label{tab:router_architecture}
\begin{tabular}{lc}
\toprule
\textbf{Component} & \textbf{Configuration} \\
\midrule
Input feature & 80-dimensional log-Mel spectrogram \\
Sample rate & 16 kHz \\
Maximum duration & 30 s \\
Frontend & Lightweight 1D convolutional frontend \\
Temporal downsampling & $2\times$ \\
Hidden dimension & 128 \\
Transformer layers & 1 \\
Attention heads & 4 \\
Feed-forward dimension & 256 \\
Pooling & Attention pooling \\
Classifier & Linear binary head \\
Output labels & clean / degraded \\
\bottomrule
\end{tabular}
\end{table}

\subsection{Router Training Data}

The router is trained with binary supervision. Clean speech is labeled as
$0$ and routed to the original Qwen3-ASR backbone, while degraded speech is
labeled as $1$ and routed to the \textsc{Mega-ASR} LoRA branch. The clean subset is
constructed from LibriSpeech, AISHELL-1, CommonVoice22, and WenetSpeech. The
degraded subset is constructed from  \textsc{Voices-in-the-wild-2M} .

The final router dataset contains 552{,}651 clean samples and 674{,}107 degraded
samples. We split the data into training, validation, and test sets, containing
1{,}104{,}084, 61{,}337, and 61{,}337 samples respectively.

\begin{table}[t]
\centering
\small
\caption{Dataset used for training the environment-aware router. Clean samples are labeled as 0 and degraded samples are labeled as 1.}
\label{tab:router_dataset}
\begin{tabular}{lcc}
\toprule
\textbf{Subset} & \textbf{Source} & \textbf{Number of samples} \\
\midrule
Clean & LibriSpeech, AISHELL-1, CommonVoice22, WenetSpeech & 552{,}651 \\
Degraded &  \textsc{Voices-in-the-wild-2M}  & 674{,}107 \\
\midrule
Train split & Mixed clean/degraded & 1{,}104{,}084 \\
Validation split & Mixed clean/degraded & 61{,}337 \\
Test split & Mixed clean/degraded & 61{,}337 \\
\bottomrule
\end{tabular}
\end{table}

On the held-out development set, the router achieves over 99.5\% binary
classification accuracy, indicating that the acoustic difference between clean
and degraded inputs can be reliably detected by a lightweight model.

\subsection{Training Objective and Optimization}

The router is trained as a binary classifier. Given an input utterance $x$ and
a label $y\in\{0,1\}$, where $y=1$ denotes degraded speech, the router predicts
$p_\theta(y\mid x)$. We optimize the standard cross-entropy loss:
\[
\mathcal{L}_{\mathrm{router}}
=
-\log p_\theta(y\mid x).
\]

During training, each audio file is resampled to 16 kHz, converted to mono, and
truncated to at most 30 seconds. We extract log-Mel spectrogram features and pad
variable-length sequences within each mini-batch. For training samples, we apply
lightweight augmentation including random gain perturbation and weak additive
noise. The model is optimized with AdamW, a warmup cosine learning-rate schedule,
gradient clipping, label smoothing, and mixed-precision training.

\begin{table}[t]
\centering
\small
\caption{Router training configuration.}
\label{tab:router_training_config}
\begin{tabular}{lc}
\toprule
\textbf{Item} & \textbf{Setting} \\
\midrule
Task & Binary clean/degraded classification \\
Input feature & Log-Mel spectrogram \\
Sample rate & 16 kHz \\
Maximum duration & 30 s \\
Loss & Cross entropy \\
Label smoothing & 0.1 \\
Optimizer & AdamW \\
Learning-rate schedule & Warmup + cosine decay \\
Warmup ratio & 0.1 \\
Gradient clipping & 1.0 \\
Mixed precision & Enabled \\
Validation metrics & Accuracy, precision, recall, F1, AUC \\
Best checkpoint criterion & Validation accuracy / AUC \\
Development accuracy & $>99.5\%$ \\
\bottomrule
\end{tabular}
\end{table}

\subsection{Integration with Qwen3-ASR and LoRA Delta Switching}

We integrate the router with Qwen3-ASR-1.7B using a single-model delta-switching
design. Instead of loading separate base and robust ASR models, we load one
Qwen3-ASR-1.7B instance and precompute the LoRA weight deltas of the robust
adapters. At inference time, the router predicts whether the input is degraded.
If the input is predicted as clean, the model keeps or switches to the base
weights; if it is predicted as degraded, the LoRA deltas are activated and the
utterance is decoded with the robust \textsc{Mega-ASR} branch.

Concretely, the system first runs the audio-quality predictor and obtains a
dirty probability $p_{\mathrm{dirty}}$. With threshold $\gamma=0.5$, routing is
defined as
\[
\mathrm{route}(x)=
\begin{cases}
\text{Mega-ASR LoRA branch}, & p_{\mathrm{dirty}}(x)\geq \gamma,\\
\text{Qwen3-ASR base branch}, & p_{\mathrm{dirty}}(x)<\gamma.
\end{cases}
\]
The LoRA switch is implemented by adding or subtracting the precomputed LoRA
delta tensors from the corresponding base weights. Therefore, switching does
not require reloading the full model and introduces only a small runtime
overhead.

\begin{algorithm}[t]
\caption{Router-guided LoRA delta switching for Qwen3-ASR}
\label{alg:router_delta_switch}
\begin{algorithmic}[1]
\Require Input audio $x$, router $g$, threshold $\gamma$
\Require Qwen3-ASR base model with preloaded LoRA deltas
\State Compute dirty probability $p_{\mathrm{dirty}} \leftarrow g(x)$
\If{$p_{\mathrm{dirty}} \geq \gamma$}
    \State Set LoRA state to active
    \State Decode $x$ with the \textsc{Mega-ASR} branch
\Else
    \State Set LoRA state to inactive
    \State Decode $x$ with the Qwen3-ASR base branch
\EndIf
\State Return transcription
\end{algorithmic}
\end{algorithm}

\subsection{Inference Overhead}

Because the router is a small single-layer classifier and LoRA switching is
implemented by adding or subtracting precomputed delta tensors, the additional
runtime cost is negligible. The router is executed once before transcription,
and the ASR model itself is not reloaded during switching. In batch inference,
we first group utterances by routing decision and then decode the LoRA-routed
and base-routed groups separately, further reducing unnecessary switching.

We measure inference time on CHiME-4 using the same evaluation pipeline for
direct Qwen3-ASR inference and router-guided inference. As shown in
Table~\ref{tab:router_latency}, the routed system has a total runtime of
371 seconds, compared with 374 seconds for direct Qwen3-ASR inference. The
relative difference is $-0.8\%$, which is within normal runtime fluctuation.
Therefore, the router and delta-switching mechanism introduce no measurable
inference overhead in practice.

\begin{table}[t]
\centering
\small
\caption{Inference-time overhead of router-guided LoRA switching on CHiME-4. The routed system shows comparable runtime to direct Qwen3-ASR inference, with relative difference below 1\%.}
\label{tab:router_latency}
\begin{tabular}{lccc}
\toprule
\textbf{System} & \textbf{Dataset} & \textbf{Total runtime} & \textbf{Relative difference} \\
\midrule
Qwen3-ASR-1.7B
& CHiME-4 & 374 s & -- \\

Qwen3-ASR + router + LoRA delta switch
& CHiME-4 & 371 s & $-0.8\%$ \\
\bottomrule
\end{tabular}
\end{table}

\newpage
\section{Training and Implementation Details}
\subsection{A2S-SFT Hyperparameters}
\label{app:a2s_sft_hparams}

This section reports the training configuration of the Acoustic-to-Semantic
Supervised Fine-Tuning (A2S-SFT) stage. A2S-SFT is implemented as a three-phase
training procedure: \textit{(i)} encoder-aligner acoustic adaptation,
\textit{(ii)} LLM-side semantic adaptation, and \textit{(iii)} joint
acoustic-semantic adaptation. All phases are initialized from Qwen3-ASR-1.7B and
use LoRA-based parameter-efficient fine-tuning. Unless otherwise specified, the
effective batch size is set to $128$.

\paragraph{Training schedule.}
The three phases differ in both trainable scope and data schedule. In the first
phase, only the acoustic encoder and the speech-to-LLM aligner are updated. This
phase is the only stage where we apply a WER-graded curriculum. Specifically, the
training subset is progressively expanded from $\mathrm{WER}<30\%$ to
$\mathrm{WER}<50\%$, and finally to $\mathrm{WER}<70\%$. This schedule provides a
stable acoustic warm start before exposing the encoder-aligner stack to harder
and noisier samples.

The second and third phases do not use the progressive WER curriculum. Instead,
they are trained directly on the full targeted split. In Phase II, the acoustic
encoder and aligner are frozen, and only the LLM-side LoRA parameters are updated
to adapt the language model to noisy transcription recovery. In Phase III, the
encoder, aligner, and LLM are jointly updated with LoRA to align the acoustic
representations and semantic decoding behavior end-to-end.

\begin{figure}[t]
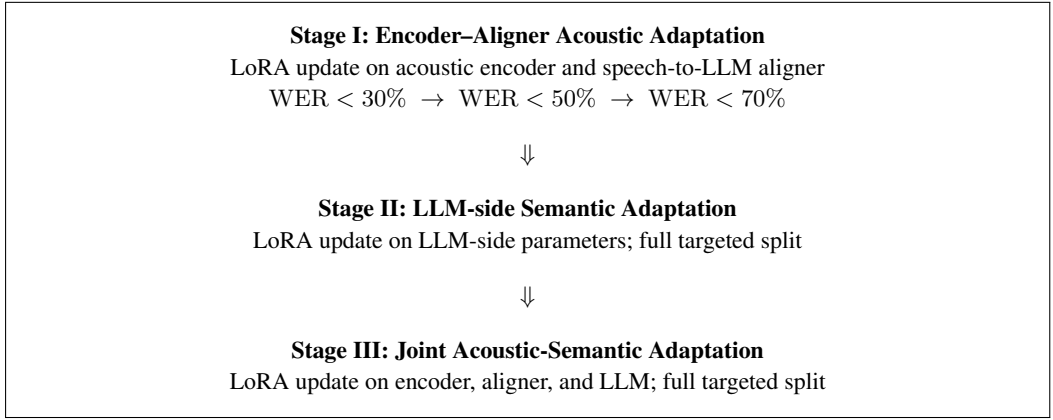

\centering
\setlength{\fboxsep}{7pt}
\fbox{
\begin{minipage}{0.94\linewidth}
\centering
\small
\renewcommand{\arraystretch}{1.18}
\begin{tabular}{c}
\textbf{Stage I: Encoder--Aligner Acoustic Adaptation} \\
LoRA update on acoustic encoder and speech-to-LLM aligner \\
$\mathrm{WER}<30\% \;\rightarrow\; \mathrm{WER}<50\% \;\rightarrow\; \mathrm{WER}<70\%$ \\
\\[-1mm]
$\Downarrow$ \\
\\[-1mm]
\textbf{Stage II: LLM-side Semantic Adaptation} \\
LoRA update on LLM-side parameters; full targeted split \\
\\[-1mm]
$\Downarrow$ \\
\\[-1mm]
\textbf{Stage III: Joint Acoustic-Semantic Adaptation} \\
LoRA update on encoder, aligner, and LLM; full targeted split \\
\end{tabular}
\end{minipage}
}
\vspace{-1mm}
\caption{
A2S-SFT training schedule. The WER-graded curriculum is applied only in Stage I
for encoder-aligner adaptation. Stages II and III are trained on the full
targeted split.
}
\label{fig:a2s_sft_pipeline}
\vspace{-2mm}
\end{figure}

\paragraph{Data construction.}
For the encoder-aligner warm-start phase, we sample $30$K utterances from the
training pool and use them to construct the WER-graded acoustic curriculum. The
curriculum first uses relatively reliable samples to stabilize the acoustic
interface, then gradually introduces more challenging samples with higher WER.
The subsequent LLM-side and joint adaptation phases use the full targeted split
constructed from the same preprocessing pipeline. The validation set is kept
disjoint from the training set and is used for checkpoint monitoring and failure
pattern inspection.

\paragraph{Stage-wise hyperparameters.}
Table~\ref{tab:a2s_sft_stage_hparams} summarizes the main hyperparameters of the
three phases. Phase I adapts the acoustic tower and the projection/aligner module
with LoRA. In our implementation, the trainable acoustic scope focuses on the
upper acoustic blocks and the projection module, so that the model can adjust
high-level acoustic representations while keeping the majority of the pretrained
backbone stable. Phase II freezes the acoustic side and updates the LLM-side
LoRA parameters. Phase III jointly updates all three module groups, with a smaller
learning rate for the acoustic encoder and aligner to avoid disrupting the
representations obtained in Phase I.

\begin{table}[t]
\centering
\small
\caption{
Stage-wise hyperparameters of A2S-SFT. The WER curriculum is used only in
Phase I; Phases II and III are trained on the full targeted split.
}
\label{tab:a2s_sft_stage_hparams}
\renewcommand{\arraystretch}{1.13}
\begin{tabular}{p{0.20\linewidth} p{0.24\linewidth} p{0.24\linewidth} p{0.24\linewidth}}
\toprule
\textbf{Setting}
& \textbf{Phase I}
& \textbf{Phase II}
& \textbf{Phase III} \\
\midrule

Training role
& Acoustic warm start
& Semantic adaptation
& Joint alignment \\

Trainable modules
& Encoder + aligner
& LLM
& Encoder + aligner + LLM \\

Data schedule
& WER-graded curriculum
& Full targeted split
& Full targeted split \\

WER range
& $<30\% \rightarrow <50\% \rightarrow <70\%$
& Full targeted range
& Full targeted range \\

Per-device batch size
& 8
& 8
& 8 \\

Number of GPUs
& 2
& 2
& 2 \\

Gradient accumulation
& 8
& 8
& 8 \\

Effective batch size
& 128
& 128
& 128 \\

Epochs
& 2
& 1
& 1 \\

Encoder learning rate
& $1.0 \times 10^{-6}$
& frozen
& $5.0 \times 10^{-7}$ \\

Aligner learning rate
& $1.0 \times 10^{-6}$
& frozen
& $5.0 \times 10^{-7}$ \\

LLM learning rate
& frozen
& $1.0 \times 10^{-6}$
& $1.0 \times 10^{-6}$ \\

Warmup ratio
& 0.05
& 0.05
& 0.03 \\

Weight decay
& 0.01
& 0.01
& 0.01 \\

Maximum gradient norm
& 1.0
& 1.0
& 1.0 \\

LoRA rank $r$
& 8
& 8
& 8 \\

LoRA alpha
& 16
& 16
& 16 \\

LoRA dropout
& 0.05
& 0.05
& 0.05 \\

Checkpoint interval
& 200 steps
& 200 steps
& 200 steps \\

Saved weights
& Adapter only
& Adapter only
& Adapter / merged adapter \\
\bottomrule
\end{tabular}
\vspace{-2mm}
\end{table}

\paragraph{Optimization details.}
All phases are trained with distributed data parallelism on two GPUs. The
effective batch size is computed as
\[
B_{\mathrm{eff}}
=
B_{\mathrm{device}}
\times N_{\mathrm{GPU}}
\times N_{\mathrm{accum}}
=
8 \times 2 \times 8
=
128.
\]
We use conservative learning rates because the model is initialized from a
pretrained ASR-LLM checkpoint rather than trained from scratch. In the final joint
phase, the encoder and aligner learning rates are reduced to
$5.0\times10^{-7}$, while the LLM-side learning rate remains
$1.0\times10^{-6}$. This asymmetric setting helps preserve the acoustic
adaptation from Phase I while still allowing the language model to adjust to the
full noisy transcription distribution. Gradients are clipped to $1.0$ in all
phases, and checkpoints are saved every $200$ optimization steps. We save
adapter-only checkpoints during intermediate phases to reduce storage overhead
and simplify later merging.

\begin{table}[t]
\centering
\small
\caption{
Common implementation settings used in A2S-SFT.
}
\label{tab:a2s_sft_common_settings}
\renewcommand{\arraystretch}{1.12}
\begin{tabular}{p{0.34\linewidth} p{0.50\linewidth}}
\toprule
\textbf{Item} & \textbf{Setting} \\
\midrule
Backbone model
& Qwen3-ASR-1.7B \\

Training type
& LoRA-based supervised fine-tuning \\

Distributed training
& 2-GPU training with one process per GPU \\

Per-device batch size
& 8 \\

Gradient accumulation steps
& 8 \\

Effective batch size
& 128 \\

Optimizer regularization
& Weight decay 0.01 \\

Gradient clipping
& Maximum gradient norm 1.0 \\

Warmup
& Linear warmup with ratio 0.05 or 0.03 depending on phase \\

Checkpointing
& Save every 200 optimization steps \\

Checkpoint format
& Adapter-only during intermediate phases; merged adapter for downstream use \\

Model selection
& Validation WER together with inspection of empty, hallucinated, and off-audio
outputs \\
\bottomrule
\end{tabular}
\vspace{-2mm}
\end{table}

\paragraph{Preliminary training variants.}
Before fixing the above schedule, we examined several alternative update orders.
These comparisons were used to validate the need for staged optimization rather
than as separate model variants in the final system. Directly training all modules
from the beginning was less stable on medium- and high-WER samples, since the
language model could adapt to unreliable acoustic representations before the
encoder-aligner interface became sufficiently grounded. Training only the
encoder-aligner improved acoustic consistency but gave limited gains on heavily
corrupted samples requiring semantic recovery. Conversely, adapting the LLM before
the acoustic warm start made the model more prone to relying on language priors.
The final schedule therefore uses encoder-aligner adaptation first, LLM-side
adaptation second, and joint acoustic-semantic alignment last.

\begin{table}[t]
\centering
\small
\caption{
Preliminary A2S-SFT variants considered during development.
}
\label{tab:a2s_sft_variants}
\renewcommand{\arraystretch}{1.12}
\begin{tabular}{p{0.30\linewidth} p{0.58\linewidth}}
\toprule
\textbf{Variant} & \textbf{Observation} \\
\midrule

Direct joint SFT from the beginning
& Less stable in medium- and high-WER regimes; the model could adapt to
unreliable acoustic representations early in training. \\

Encoder-aligner only
& Improved acoustic grounding, but provided limited recovery when the acoustic
evidence was incomplete or severely corrupted. \\

LLM adaptation before acoustic warm start
& Increased reliance on the language prior before the acoustic interface was
sufficiently stabilized. \\

No WER curriculum in Phase I
& Produced larger validation fluctuations during the acoustic warm-start stage. \\

Final three-phase schedule
& Provided the most stable training behavior by separating acoustic adaptation,
semantic adaptation, and final end-to-end alignment. \\
\bottomrule
\end{tabular}
\vspace{-2mm}
\end{table}

\subsection{DG-WGPO Hyperparameters}
\label{app:dg_wgpo_hparams}

This section provides the implementation and training details of
Dual-Granularity WER-Gated Policy Optimization (DG-WGPO). DG-WGPO is implemented
with DAPO-style policy optimization in an RLHF framework. Since Qwen3-ASR is not a
standard text-only causal language model, we introduce a custom multimodal
adaptation layer to make the audio encoder, speech-to-language aligner, and
language model compatible with group-based policy optimization.

\paragraph{Framework and model adaptation.}
Qwen3-ASR takes both an audio signal and a text-side prompt as input. Therefore,
directly treating it as a pure text model would break the rollout and loss
construction used in GRPO/DAPO. We adapt Qwen3-ASR as a multimodal policy model
while keeping its official inference behavior unchanged. The adaptation mainly
addresses four issues: \textit{(i)} preserving the original audio preprocessing
and prompt construction protocol, \textit{(ii)} exposing a training interface that
accepts both text tokens and acoustic features, \textit{(iii)} making the inner
language model compatible with LoRA-based policy updates, and \textit{(iv)}
ensuring that rollout completions retain the raw ASR format for reward parsing.

\begin{table}[t]
\centering
\small
\caption{
Summary of the Qwen3-ASR adaptation used for DG-WGPO. We only list the
model-level adaptation principles and omit implementation-specific function or
class names.
}
\label{tab:dg_wgpo_asr_adaptation}
\begin{tabular}{p{0.30\linewidth} p{0.60\linewidth}}
\toprule
\textbf{Adaptation item} & \textbf{Purpose} \\
\midrule
Multimodal model loading
& Load Qwen3-ASR as an audio-language policy model rather than a text-only
decoder. \\

Official processor consistency
& Keep the same audio preprocessing, tokenizer behavior, and prompt protocol
between inference, rollout, and RL training. \\

Multimodal forward interface
& Allow the trainer to pass text tokens, text masks, acoustic features, acoustic
masks, and response labels in a unified training call. \\

Language-model interface alignment
& Expose the inner language-model embeddings and output head to the LoRA/RLHF
trainer without changing the Qwen3-ASR architecture. \\

Module grouping
& Separate the model into language model, acoustic encoder, and aligner groups,
so that update scopes can be controlled explicitly. \\

Rollout re-encoding
& Re-encode sampled completions together with the original multimodal prompt and
apply loss only on the generated response tokens. \\

Padding policy
& Use left padding for text-side sequences and temporal padding for audio-side
features, matching decoder-only generation and acoustic feature batching. \\

Raw-output preservation
& Preserve the original ASR completion format during decoding, allowing the
reward function to parse empty outputs, language prefixes, repetitions, and
format irregularities consistently. \\
\bottomrule
\end{tabular}
\vspace{-2mm}
\end{table}

\paragraph{Data and initialization.}
The DG-WGPO stage is initialized from the A2S-SFT LoRA-merged checkpoint rather
than from the original Qwen3-ASR checkpoint. This ensures that the initial policy
already has stable acoustic grounding and reasonable transcription ability before
entering reinforcement learning. The RL training and validation sets are
constructed as targeted WER-aware splits. Unlike general SFT data, the RL data are
enriched with medium- and high-WER examples, while relatively clean utterances are
reduced to prevent the policy update from being dominated by easy samples. This
design matches the goal of DG-WGPO: improving robustness in the regimes where
standard supervised fine-tuning and WER-only rewards provide limited corrective
signals.

Each RL example contains a multimodal prompt, an audio input, a reference
transcription, the base-model prediction, and the base WER used for data
selection and analysis. Table~\ref{tab:dg_wgpo_data_schema} summarizes the data
schema. During training, the reference transcription is used only by the reward
function; the policy is optimized from sampled completions and their group-wise
relative rewards.

\begin{table}[t]
\centering
\small
\caption{
JSONL data schema used in DG-WGPO. The absolute audio paths are omitted from the
paper.
}
\label{tab:dg_wgpo_data_schema}
\begin{tabular}{p{0.22\linewidth} p{0.68\linewidth}}
\toprule
\textbf{Field} & \textbf{Description} \\
\midrule
\texttt{messages}
& System and user instructions that define the ASR task, e.g., transcribe the
given audio and output plain text only. \\

\texttt{audios}
& A list containing the audio file associated with the current prompt. Each
training example uses one audio input. \\

\texttt{solution}
& Reference transcription used for WER computation and reward evaluation. \\

\texttt{prediction}
& Transcription generated by the initialization model. This field is used for
data targeting and diagnostic comparison, not as a supervised label. \\

\texttt{base\_wer}
& WER of the initialization model on the current sample. We use it to emphasize
medium- and high-WER regions in RL data construction and analysis. \\

\texttt{meta}
& Optional metadata field for bookkeeping. \\
\bottomrule
\end{tabular}
\vspace{-2mm}
\end{table}

\paragraph{Policy optimization setup.}
DG-WGPO uses GRPO-style group-relative advantage estimation with the DAPO loss.
The reported main run uses three GPUs, and the same setting can be scaled to four
or eight GPUs by increasing the number of distributed processes. We keep the
per-device batch size, number of generations, and reward settings unchanged when
scaling the number of GPUs.

In the main run, the language model, acoustic encoder, and aligner are all allowed
to receive LoRA updates. Although all three module groups participate in policy
optimization, the total number of trainable parameters remains controlled because
the update is parameter-efficient. This full-scope LoRA update is important for
DG-WGPO because the reward simultaneously targets acoustic grounding and semantic
reconstruction. Updating only the language model improves language-side fluency
but is less effective for acoustically induced substitutions and omissions, while
updating only the encoder-aligner limits sentence-level recovery in high-WER
cases. Therefore, the reported DG-WGPO results use LoRA updates across the
acoustic encoder, aligner, and LLM.

\begin{table}[t]
\centering
\small
\caption{
Main training hyperparameters of DG-WGPO. The effective prompt batch size shown
below corresponds to the three-GPU main run.
}
\label{tab:dg_wgpo_train_hparams}
\begin{tabular}{p{0.38\linewidth} p{0.46\linewidth}}
\toprule
\textbf{Hyperparameter} & \textbf{Setting} \\
\midrule
Initialization
& A2S-SFT LoRA-merged Qwen3-ASR-1.7B checkpoint \\

Optimization method
& GRPO-style advantage estimation with DAPO loss \\

Training type
& LoRA \\

Trainable scope
& Acoustic encoder + aligner + language model \\

Main number of GPUs
& 3 \\

Scalable GPU settings
& 3 / 4 / 8 GPUs \\

Per-device train batch size
& 4 \\

Per-device evaluation batch size
& 4 \\

Gradient accumulation steps
& 16 \\

Effective prompt batch size
& $4 \times 3 \times 16 = 192$ \\

Number of generations per prompt
& 12 \\

Evaluation generations per prompt
& 4 \\

Maximum completion length
& 256 tokens \\

Learning rate
& $5.0 \times 10^{-5}$ \\

Learning-rate scheduler
& Cosine decay \\

Warmup ratio
& 0.03 \\

KL coefficient $\beta$
& 0.04 \\

DAPO upper clipping parameter
& 0.28 \\

Number of RL iterations
& 2 \\

Dynamic sampling
& Enabled \\

Maximum resampling times
& 4 \\

Overlong filtering
& Enabled \\

Truncation strategy
& Delete overlong samples \\

Checkpoint interval
& Every 20 steps \\

Logging interval
& Every 5 steps \\
\bottomrule
\end{tabular}
\vspace{-2mm}
\end{table}

\paragraph{Rollout and generation protocol.}
For each prompt, DG-WGPO samples $K=12$ candidate transcriptions and computes
group-relative rewards. We use stochastic decoding because the policy update
requires sufficient intra-group diversity: if all completions are nearly
identical, the advantage signal collapses. However, excessive exploration can
increase hallucinations, overlong outputs, and format violations. We therefore
choose the generation parameters through a small exploratory probing stage rather
than fixing them heuristically.

Let $b_i$ denote the WER of the initialization model on example $i$, and let
$H_{i,k}^{(T)}$ be the $k$-th sampled completion under temperature $T$. We use two
statistics to compare temperature settings. The first measures whether the sample
group contains a potentially better candidate:
\begin{equation}
\mathrm{CPI}_{\delta}(T)
=
\frac{1}{N}
\sum_{i=1}^{N}
\mathbb{I}
\left[
\min_{1 \leq k \leq K}
\mathrm{WER}\!\left(H_{i,k}^{(T)}, R_i\right)
\leq b_i - \delta
\right],
\end{equation}
where $\mathrm{CPI}$ denotes the candidate potential indicator. The second
measures whether the reward-selected candidate preserves the base transcription
ability:
\begin{equation}
\mathrm{BAP}_{\delta}(T)
=
\frac{1}{N}
\sum_{i=1}^{N}
\mathbb{I}
\left[
\mathrm{WER}\!\left(H_{i,k^\star}^{(T)}, R_i\right)
\leq b_i + \delta
\right],
\quad
k^\star = \arg\max_{k} R\!\left(H_{i,k}^{(T)}, R_i\right),
\end{equation}
where $\mathrm{BAP}$ denotes base-ability preservation. In practice, we also
monitor the valid-output rate, including non-empty outputs, non-repetitive
outputs, and completions within the maximum length. The final temperature is
chosen to balance candidate potential, base-ability preservation, and output
validity.

\begin{table}[t]
\centering
\small
\caption{
Generation settings used in the main DG-WGPO run.
}
\label{tab:dg_wgpo_generation_hparams}
\begin{tabular}{p{0.34\linewidth} p{0.50\linewidth}}
\toprule
\textbf{Hyperparameter} & \textbf{Setting} \\
\midrule
Number of generations
& 12 \\

Evaluation generations
& 4 \\

Temperature
& 0.50 \\

Top-$p$
& 0.95 \\

Top-$k$
& 50 \\

Repetition penalty
& 1.08 \\

Maximum completion length
& 256 tokens \\

Dynamic sampling
& Enabled \\

Maximum resampling times
& 4 \\

Overlong filtering
& Enabled \\
\bottomrule
\end{tabular}
\vspace{-2mm}
\end{table}

\begin{table}[t]
\centering
\small
\caption{
Temperature probing protocol. The final run uses $T=0.50$ because it provides
moderate exploration while preserving the base ASR behavior.
}
\label{tab:dg_wgpo_temperature_probe}
\begin{tabular}{p{0.16\linewidth} p{0.34\linewidth} p{0.40\linewidth}}
\toprule
\textbf{Temperature} & \textbf{Observed tendency} & \textbf{Role in selection} \\
\midrule
0.30
& Conservative decoding with low intra-group diversity
& Used to check the lower-exploration regime; often yields weak advantage
dispersion. \\

0.50
& Moderate diversity with stable ASR formatting and relatively high valid-output
rate
& Selected as the default setting by balancing candidate potential and
base-ability preservation. \\

0.70
& Higher diversity and more possible corrections
& Used to probe whether more aggressive sampling reveals better candidates, but
requires stronger filtering. \\

0.90
& Strong exploration with higher risk of hallucination, off-audio text, and
overlong outputs
& Used only as a stress test for reward robustness and filtering behavior. \\
\bottomrule
\end{tabular}
\vspace{-2mm}
\end{table}

\paragraph{Reward tuning and diagnostics.}
The main reward function follows the DG-WGPO formulation described in the main
text. We use a WER-gated dynamic reward with $\tau=0.5$, soft-error discount
$\alpha_s=0.4$, and dynamic-reward weight $\alpha_{\mathrm{dyn}}=0.6$. For samples
below the WER gate, the reward emphasizes token-level refinement; for samples
above the gate, it assigns more weight to sentence-level structural recovery.
This design is especially important for the targeted RL split, where many
examples contain medium- or high-WER predictions and the standard WER reward can
become less discriminative.

\begin{table}[t]
\centering
\small
\caption{
Reward hyperparameters used in DG-WGPO.
}
\label{tab:dg_wgpo_reward_hparams}
\begin{tabular}{p{0.38\linewidth} p{0.46\linewidth}}
\toprule
\textbf{Reward hyperparameter} & \textbf{Setting} \\
\midrule
Static WER reward
& $1-\mathrm{WER}$ \\

Repetition gate
& Enabled \\

Soft substitution threshold
& Character-level edit similarity $\geq 0.5$ \\

Soft-error discount $\alpha_s$
& 0.4 \\

WER gate threshold $\tau$
& 0.5 \\

Dynamic reward weight $\alpha_{\mathrm{dyn}}$
& 0.6 \\

Low-WER fusion
& $0.75 R_{\mathrm{fine}} + 0.25 R_{\mathrm{struc}}$ \\

High-WER fusion
& $0.25 R_{\mathrm{fine}} + 0.75 R_{\mathrm{struc}}$ \\

Reward scaling
& Group-wise scaling \\

Length and overlong control
& Enabled through rollout filtering and reward diagnostics \\
\bottomrule
\end{tabular}
\vspace{-2mm}
\end{table}

We tune the reward by inspecting rollout groups rather than relying only on the
scalar training reward. For each diagnostic group, we compare the reference, the
initial model prediction, sampled hypotheses, component rewards, and final reward
ranking. This allows us to check whether a higher reward corresponds to a real
error reduction, such as correcting acoustically plausible substitutions,
recovering omitted content, reducing repeated phrases, or improving sentence-level
structure. We also inspect failure cases where the scalar reward prefers a
shorter but incomplete hypothesis, a fluent but off-audio hypothesis, or a
format-valid but semantically incorrect transcription. These diagnostics are used
to calibrate the WER gate, the soft-error discount, and the balance between the
static and dynamic rewards.

Model selection is based on validation WER together with rollout quality
statistics, including empty-output rate, repetition rate, overlong-output rate,
and the behavior of reward-selected candidates on medium- and high-WER samples.
This avoids selecting checkpoints that over-optimize a single reward component
while degrading transcription faithfulness.







\section{Additional Related works}
\paragraph{Traditional Robust ASR Methods.}
Robust automatic speech recognition has been widely studied to improve transcription under noise, reverberation, channel mismatch, speaker variation, and domain shift. Traditional methods usually rely on speech enhancement, feature normalization, speaker adaptation, multi-condition training, and language-model rescoring. With the development of end-to-end ASR, data augmentation and large-scale pretraining have become dominant solutions. Representative works include SpecAugment~\cite{park2019specaugment}, Conformer~\cite{conformer}, wav2vec 2.0~\cite{wav2vec}, HuBERT~\cite{hubert}, WavLM~\cite{wavllm}, and Whisper~\cite{whisper}. These methods greatly improve robustness, but they are still mainly optimized for transcription accuracy and are usually evaluated by word error rate, rather than semantic understanding or reasoning over speech.

\paragraph{Large Audio Language Models.}
Large Audio Language Models (LALMs) connect speech or general audio signals with large language models, enabling audio-conditioned instruction following, question answering, and reasoning. Compared with conventional ASR systems, LALMs are attractive for ASR because they can use linguistic knowledge and contextual reasoning to recover corrupted or ambiguous speech. Existing LALMs have shown strong capabilities in audio-conditioned understanding, instruction following, and speech-language reasoning~\cite{LTU,pengi,qwen2.5omni,audiopalm,wavllm,audioflamingo,mini-omni,mini-omni2,stepaudio2,seedasr}. However, this ability may also introduce hallucinations, where the model generates plausible but incorrect transcriptions that are not grounded in the input audio. Recent works have further explored reasoning-based ASR, attempting to use audio-language reasoning to improve recognition beyond direct acoustic decoding~\cite{audio-reasoner,mini-omni-reasoner,huang2025rebellion}.

\paragraph{Speech Recognition Datasets and Benchmarks.}
Speech recognition datasets and benchmarks provide the basis for evaluating ASR performance under different acoustic and linguistic conditions. Commonly used clean or read-speech datasets include LibriSpeech~\cite{librispeech} and TED-LIUM~\cite{ted-lium}, while Switchboard~\cite{switchboard} is widely used for conversational speech recognition. Common Voice~\cite{commonvoice} supports multilingual and diverse-speaker ASR evaluation. For noisy, far-field, and meeting scenarios, representative benchmarks include CHiME~\cite{CHIME4}, AMI~\cite{ami}, and Speech Robust Bench~\cite{rubostspeechbench}. These datasets mainly evaluate transcription quality with WER, making them suitable for measuring ASR robustness but less focused on reasoning or instruction-following ability.

\end{document}